\documentclass[a4paper,oneside,fleqn,12pt]{article}

\usepackage{amsmath,amssymb,amsfonts,amsthm,type1cm,bm,color,colortbl, mathrsfs}
\usepackage{booktabs, geometry}
\usepackage{natbib}
\usepackage{rotating}
\usepackage{authblk}
\usepackage{accents}
\usepackage{yfonts}
\usepackage{pdflscape}
\usepackage{setspace}
\usepackage{longtable}
\usepackage{bigstrut}
\usepackage{mathtools}
\usepackage{enumitem}
\usepackage[dvipsnames]{xcolor}
\usepackage{comment}
\usepackage{float}
\usepackage{multibib}

\usepackage{graphicx}
\RequirePackage[colorlinks,citecolor=blue,urlcolor=blue,hyperfootnotes=false]{hyperref}

\newcites{suppl}{References for Supplementary Material}
\mathtoolsset{showonlyrefs=true}
\DeclareMathOperator\E{\mathbb{E}}

\DeclareMathOperator\Pro{\mathbb{P}}

\DeclareMathOperator\sgn{sgn}

\def\cF{\mathcal{F}}

\def\cS{\mathcal{S}}

\def\bbR{\mathbb{R}}

\def\bbN{\mathbb{N}}

\def\what{\widehat}

\newlength{\dhatheight}

\newtheorem{thm}{Theorem}
\newtheorem{con}{Condition}
\newtheorem{rem}{Remark}
\newtheorem{ex}{Example}
\newtheorem{proc}{Procedure}
\newtheorem{proper}{Property}

\makeatletter
\def\section{\@startsection {section}{1}{\z@}{-3.5ex plus -1ex minus-.2ex}{2.3ex plus .2ex}{\large\bf}}
\makeatother

\makeatletter
\def\subsection{\@startsection {subsection}{1}{\z@}{-3.5ex plus -1ex minus-.2ex}{2.3ex plus .2ex}{\normalsize\bf}}
\makeatother

\title{\textbf{Sequential Correct Screening and \\Post-Screening Inference}}

\author[$*$]{\textsc{Masaki Toyoda}
}
\author[$\dagger$]{\textsc{Yoshimasa Uematsu}
}

\affil[*]{\textit{Department of Economics, Hitotsubashi University}}
\affil[$\dagger$]{\textit{Department of Social Data Science, Hitotsubashi University}}

\begin{document}

\renewcommand{\theequation}{\thesection.\arabic{equation}}
\makeatletter
\@addtoreset{equation}{section}
\makeatother

\maketitle

\begin{abstract}
Selecting the top‑$m$ variables with the $m$ largest population parameters from a larger set of candidates is a fundamental problem in statistics. In this paper, we propose a novel methodology called Sequential Correct Screening (SCS), which sequentially screens out variables that are not among the top‑$m$. A key feature of our method is its anytime validity; it provides a sequence of variable subsets that, with high probability, always contain the true top‑$m$ variables. Furthermore, we develop a post-screening inference (PSI) procedure to construct confidence intervals for the selected parameters. Importantly, this procedure is designed to control the false coverage rate (FCR) whenever it is conducted---an aspect that has been largely overlooked in the existing literature. We establish theoretical guarantees for both SCS and PSI, and demonstrate their performance through simulation studies and an application to a real-world dataset on suicide rates.
\end{abstract}

\section{Introduction}\label{sec:intro}
In the era of big data, \textit{exploratory data analysis} (EDA) (\citealp{tukey1962future})---which aims to discover insights or generate hypotheses from data to contribute to science---has become increasingly essential. One important EDA task is to identify the top-$m$ variables with the largest population parameters (e.g., means) among $k \ge m$ candidates. This problem arises in many domains: in public health, for instance, identifying countries with the highest suicide rates is informative; in marketing, selecting advertisements with the highest click-through rates is crucial. Such problems are broadly known as \textit{ranking and selection} (R\&S) problems in the statistics literature. Finding the single best variable, in particular, is also referred to as the \textit{best arm identification} (BAI) problem, closely related to the \textit{multi-armed bandit} (MAB) literature. Seminal contributions include \citet{bechhofer1954single} for R\&S and \citet{robbins1952some} for BAI. Research in this area remains active today.

\subsection{Sequential correct screening and post-screening inference}

In this paper, we propose a novel methodology within the R\&S literature called \textit{Sequential Correct Screening} (SCS). While the primary goal, as with many existing approaches, is to identify the top-$m$ variables, our method is distinguished by two key innovations.

First, our framework provides a variable subset that includes the true top-$m$ variables with high probability at any sample size and dimensionality. This is achieved by \textit{sequentially} screening out variables unlikely to belong to the top-$m$. Such \textit{anytime validity} is particularly advantageous in applications requiring ongoing monitoring or interim reporting. In contrast, many existing methods are designed to output a size $m$ subset only after the sufficient evidence accumulates to verify the top-$m$. Consequently, these methods cannot make any decision until the experiment concludes and often necessitate a large sample size to draw conclusions. We will review the literature in Section \ref{sec:related}.

Second, our methodology enables the construction of confidence intervals (CIs) for the selected variables to control the \textit{false coverage rate} (FCR) below a pre-specified level. The FCR is an error notion introduced by \cite{benjamini2005false}, defined as the rate of false coverages of intervals constructed for a random subset of selected parameters. We call this procedure the \textit{post-screening inference} (PSI). Importantly, it is designed to manage the anytime validity of SCS. 
A major limitation of existing works is that they have focused primarily on identifying the top-$m$, with little to no study on inference for these selected variables, which inherently restricts their applicability.

Following the methodological development, we provide both theoretical and empirical validation for these two useful properties through simulations and real-data examples. Theoretically, we prove that under reasonable conditions, correct screening is achieved, and the top-$m$ variables are exactly identified eventually. Notably, the sequential property (anytime validity) is established by Ville's inequality for nonnegative supermartingales, representing a novel and flexible approach in this context. For post-screening inference, by carefully accounting for the bias due to screening and multiplicity, we achieve the FCR control for the CIs below a pre-specified level. Furthermore, through simulations, we demonstrate the performance of our method in comparison with existing method. Finally, our method is applied to the suicide rates data to find the countries with highest suicide rates.

\subsection{Comparison to related works}\label{sec:related}

Our methods, SCS and PSI, are included in the R\&S/BAI and selective inference literature, respectively. We discuss how our methods are positioned relative to prior works.

Many studies in R\&S and BAI \citep{bechhofer1954single, rinott1978two, auer2002finite, audibert2010best, jamieson2014lil} have primarily focused on one-shot methods, where selection is conducted only once at the end of data collection. While these methods can identify the largest parameter, they provide no output until a sufficient sample size is obtained. Notably, if the difference between the largest and second-largest parameters---referred to as the \textit{indifference zone} (IZ)---is small, the required sample size tends to be very large.

On the other hand, similar to our approach, several sequential methods have been proposed where selection is conducted each time new data become available. Instead of exactly identifying the largest, these methods sequentially output a subset by screening out parameters that appear not to be the largest. Consequently, even with a small sample size, they can provide meaningful output. We compare these sequential procedures based on four key aspects: (1) whether the method is restricted to $m=1$; (2) whether a parametric assumption is required; (3) whether the IZ must be known; and (4) whether there is a constraint on the data sampling rule. A detailed comparison is summarized in Table \ref{table:comparison} and elaborated below.

Early works such as \cite{paulson1964sequential}, \cite{swanepoel1976sequential}, and \cite{kim2001fully} are limited to $m=1$, assume a multivariate normal distribution, require a known IZ, and mandate that all unscreened variables be sampled in lockstep. \cite{fan2016indifference} relaxes the requirement for known IZ but retains the other three limitations. \cite{even2006action} further relaxes the distributional assumption to sub-Gaussian but is still constrained by $m=1$ and the sampling rule.  \cite{kalyanakrishnan2010efficient} allows for any $m$ and sub-Gaussian data but still requires a known IZ and a restricted sampling rule. Finally, the \textit{Lower and Upper Confidence Bounds} (LUCB) method by \cite{kalyanakrishnan2012pac} addresses all these points. However, it is not a screening method and thus provides no output with a small sample size. Given the background, building upon the LUCB framework, we develop the first screening method that addresses all these limitations simultaneously.

\begin{table}[H]
  \centering \footnotesize
  \caption{Comparison of methods. All methods are sequential screening type, except \cite{kalyanakrishnan2012pac}.}
  \label{table:comparison}
  \begin{tabular}{lcccc}
    \toprule
    \textbf{Method} & Any $m$ & Nonparametric & Unknown IZ & Sampling  \\
    \midrule
    \cite{paulson1964sequential} & -- & -- & -- & --  \\
    \cite{swanepoel1976sequential} & -- & -- & -- & --  \\
    \cite{kim2001fully} & -- & -- & -- & -- \\
    \cite{fan2016indifference} & -- & -- & \checkmark & --  \\
    \cite{even2006action} & -- & \checkmark & \checkmark & --  \\
    \cite{kalyanakrishnan2010efficient} & \checkmark & \checkmark & -- & --  \\
    \cite{kalyanakrishnan2012pac} & \checkmark & \checkmark & \checkmark & \checkmark  \\
    \textbf{Our proposed SCS} & \checkmark & \checkmark & \checkmark & \checkmark  \\
    \bottomrule
  \end{tabular}
\end{table}

While extensive research exists on selecting the largest parameters, inference on the selected parameters has been largely overlooked. Conversely, our PSI constructs CIs for these parameters while controlling the FCR. Thus, it accounts for both selection bias and multiplicity. Since FCR is based on the conditional distribution given the selection event, \cite{benjamini2005false} made strong distributional assumptions and restricted the selection rule to make the conditional distribution tractable. Therefore, their method is not directly applicable after our SCS. As an alternative, the e-BY approach proposed by \cite{xu2024post} allows for weaker distributional assumptions and arbitrary selection rules. The e-BY method utilizes e-variables, which are nonnegative random variables with an expectation of at most one. Our PSI is also based on the e-BY framework and controls FCR through e-variables.

\subsection{Organization}
The remainder of this paper is organized as follows. Section \ref{sec:problem} formalizes the problem. Section \ref{sec:preliminaries} introduces two sequences of statistics crucial for our procedures, and provide their examples in several situations. Section \ref{sec:proc} describes our procedures of SCS and PSI. Section \ref{sec:theory} gives their statistical theory. Section \ref{sec:simulations} demonstrates our procedures in simulation studies. Section \ref{sec:applications} applies our procedures to a real dataset. Section \ref{sec:conclusion} concludes. All the proofs are collected in Supplementary Material.

\section{Problem Statement}\label{sec:problem}
We introduce the model to be considered. After that, we describe the specific goal of SCS in Section \ref{sec:sequential} and PSI in Section \ref{sec:post-selection}.

Let $\theta_1,\dots,\theta_k\in\bbR$ be any set of unknown parameters. For example, they are a parameter of each marginal distribution $P_i$, such as a mean. 
We suppose that without loss of generality they are ordered as 
$\theta_1\geq\dots\geq\theta_k$. 
For any fixed $m\in[k]$, what we want to detect is called the \textit{$m$-promising parameters}, defined as the set of indices of the parameters, $\cS=\{i:\theta_i\geq \theta_m\}$. It becomes $\cS=[m]$ if $\theta_m>\theta_{m+1}$, but $\cS$ can be larger than $m$; if $\theta_m=\theta_{m+\ell}>\theta_{m+\ell+1}$ for some $\ell\geq 1$, it is given by $\cS=[m+\ell]$. 
A broader definition is given in \cite{kalyanakrishnan2012pac}, in which what we refer to as $m$-promising corresponds to their notion of $(0, m)$-optimal.

We define filtration $(\cF_T)_{T\in\bbN}$, where $\cF_T$ is a $\sigma$-field generated by the observed data up to time $T$. Our sampling scheme is general enough to allow irregularly spaced observations that can vary across marginals.

\subsection{First goal: Sequential correct screening}\label{sec:sequential}

For detection of the $m$-promising $\cS$, our SCS attempts to screen out parameters that are deemed sufficiently small in light of the data available up to each time $T\in\bbN$. More precisely, SCS constructs a sequence of screened sets, denoted by $(\hat\cS_T)_{T\in\bbN}$, where $\hat\cS_T$ is an $\cF_T$-measurable random subset of $[k]$ for all $T\in\bbN$ with $\hat\cS_0=[k]$, and satisfies the next property. 

\begin{proper}\normalfont\label{property}
Fix any target level $\alpha\in(0,1)$.  We have 
\begin{enumerate}
    \item[(i)] 
    $\Pro\left(\forall T \in \bbN:\hat{\cS}_T\subset\hat{\cS}_{T-1}\right)=1$; 
    \item[(ii)] $\Pro\left(\forall T \in \bbN:\cS\subset\hat{\cS}_T \right)\geq1-\alpha$;
    \item[(iii)] $\Pro\left(\hat{\cS}_T=\cS\text{ eventually}\right)\geq1-\alpha$.
\end{enumerate}
\end{proper}

Property \ref{property}(i) ensures the sequence to be a.s.\ non-increasing. Property \ref{property}(ii) further restricts $\hat{\cS}_T$ to include $\cS$ uniformly over time in probability. At last, Property \ref{property}(iii) ensures that $\hat{\cS}_T$ coincides with $\cS$ for all $T\in\bbN$ except for the first few in probability. This property (iii) excludes a trivial but meaningless construction of $(\hat{\cS}_T)_{T\in\bbN}$ satisfying the previous two properties, like $\hat{\cS}_T=[k]$ for all $T\in\bbN$. 

An important feature of Property \ref{property} is its independence from $T$, the time at which  the latest data is obtained. Thus, there is no error due to asymptotic approximation, and exact inference is possible. Furthermore, this fact also implies that we can increase the sample size $T$ as far as desired. Especially, the screening process is \textit{sequential} (\textit{anytime-valid}) in $T\in\bbN$.
This means that the desirable properties (i)--(iii) hold not only in the past, but also in the future, without requiring the sample size (i.e., the monitoring period) to be fixed in advance. This is practically attractive since whether the data are limited or sufficient up to time $T$ is reflected solely in the size of $\hat{\cS}_T$; when the amount of data is not enough, the uncertainty entails a large $\hat{\cS}_T$. If we desire a more accurate approximation of $\cS$, we just continue monitoring until sufficient data are accumulated (i.e., for large $T$). During this process, the desirable properties (i)–(iii) are guaranteed to hold at all times, $T\in\bbN$. In Section \ref{sec:proc_sequential}, we will propose constructing $(\hat{\cS}_T)_{T\in\bbN}$ that can achieve Property \ref{property}(i)--(iii).

\begin{rem}
    Lemma 1 in \cite{ramdas2020admissible} implies that Property \ref{property}(ii) is equivalent to $\Pro(\cS\subset\hat{\cS}_\tau)\geq1-\alpha$, where $\tau$ is any possibly infinite stopping time with respect to $(\cF_T)_{T\in\bbN}$. 
\end{rem}

\subsection{Second goal: Post-screening inference}\label{sec:post-selection}
Given $\hat{\cS}_T$ by SCS at a fixed time $T$, we want to construct valid CIs, $C_{iT}$, for the screened parameters, $\{\theta_i:i\in\hat{\cS}_T\}$, that control the FCR at time $T$, where
\begin{align}
    \text{FCR}_T=\E \left[\frac{|\hat{\cS}_T\cap\{i:\theta_i\notin C_{iT}\}|}{|\hat{\cS}_T|}\right],
\end{align}
to be less than or equal to $\alpha$. It should be noted that SCS can stop at any $(\cF_T)_{T\in\bbN}$-adapted stopping time $\tau$, and output $\hat{\cS}_\tau$.
For example, if the decision to stop SCS at time $T$ depends on $\hat{\cS}_T$ itself, then the time is no longer fixed but a random stopping time. In such cases, only controlling $\text{FCR}_T$ for each $T$ is not sufficient. Instead, to deal with such optional stopping of SCS, we aim to control the \textit{stopped} FCR (sFCR) of given $C_{i\tau}$:
\begin{align}
    \text{sFCR}=\E\left[\frac{|\hat{\cS}_\tau\cap\{i:\theta_i\notin C_{i\tau}\}|}{|\hat{\cS}_\tau|}\right].
\end{align}
This should be indexed by $T$ as the stopping time $\tau$ is $\cF_T$-measurable, but we omit it for notational simplicity. In general, we have $\text{FCR}_T\leq\text{sFCR}$. Thus the sFCR control implies the $\text{FCR}_T$ control, but the inverse does not hold. 

As for the target level of FCR control, while we can choose a different $\alpha$, we adopt the same one as previously introduced in SCS for notational simplicity. The modification of CIs' construction with sFCR control is important since the set of parameters for which CIs are constructed depends on the data. Making CIs in a usual way without accounting for the randomness of the screened set may result in coverage failure.

One of the simplest ways to control the sFCR is the Bonferroni correction, which constructs $C_{i\tau}^{\textsf{B}}$ such that $\Pro(\theta_i\notin C_{i\tau}^{\textsf{B}})\leq\alpha/k$ holds for each $i\in[k]$ and any $\alpha\in(0,1)$. This correction achieves the sFCR control. Indeed,  by the fact that any two random variables $V,R\in\bbN$ such that $V\leq R$ satisfy $V/(R\vee 1) \leq \mathbb{I}\{V\geq 1\}$ a.s., we have 
\begin{align}
\text{sFCR}&\leq \Pro \left(|\hat{\cS}_\tau\cap\{i:\theta_i\notin C_{i\tau}^{\textsf{B}}\}|\geq 1\right) \\
&\leq \Pro \left(|\{i:\theta_i\notin C_{i\tau}^{\textsf{B}}\}|\geq 1\right)
\leq \sum_{i\in[k]}\Pro \left(\theta_i\notin C_{i\tau}^{\textsf{B}}\right) \leq \alpha,
\end{align}
where the third inequality is due to Markov's inequality. The Bonferroni correction is very simple and always applicable; however, since it is too general to rely on the specific CIs that give the sFCR, the resulting CIs tend to be overly conservative.
In Section \ref{sec:proc_post}, we will construct narrower CIs that control the sFCR.

\section{Preliminaries}\label{sec:preliminaries}

This section introduces two sequences of statistics, which play a crucial role in our procedures in the next section. Fix an arbitrary target level $\alpha\in(0,1)$. Given data up to time $T$ or more as described in Section \ref{sec:sequential}, we require two $(\cF_T)_{T\in\bbN}$-adapted sequences of statistics, $(U_{iT}(\alpha))_{T\in\bbN}$ and $(L_{iT}(\alpha))_{T\in\bbN}$, that satisfy
\begin{align}\label{eq:bounded}
    \Pro\left(\exists T\in\bbN:\theta_i\leq L_{iT}(\alpha)\right)\leq\alpha ~~~\text{and}~~~
        \Pro\left(\exists T\in\bbN:\theta_i\geq U_{iT}(\alpha)\right)\leq\alpha
\end{align}
for each $i\in[k]$. This directly leads to the sequential aspect in Property \ref{property}. 

The construction of such $(U_{iT}(\alpha))_T$ and $(L_{iT}(\alpha))_T$ is often based on \textit{Ville's inequality} \citep{Ville1939, howard2020time}: for any \textit{nonnegative supermartingale} $(M_T)_{T\in\bbN}$ with $\E[M_1]\leq1$, we have 
\begin{align}
\Pro \left( \exists T\in\bbN:M_T\geq1/\alpha \right)\leq\alpha.
\end{align}
Consequently, once we obtain such a supermartingale adapted to $(\cF_T)_{T\in\bbN}$ as a function of $\theta_i$, inverting the inequality yields $U_{iT}(\alpha)$ and $L_{iT}(\alpha)$ that satisfy \eqref{eq:bounded}. Using such $(U_{iT}(\alpha))_T$ and $(L_{iT}(\alpha))_T$ satisfying \eqref{eq:bounded} for all $i\in[k]$ and $\alpha\in(0,1)$, we propose methodologies for SCS and PSI in Sections \ref{sec:proc_sequential} and \ref{sec:proc_post}, respectively.

\subsection{Examples}\label{sec:ex}

We give four examples of \eqref{eq:bounded}. Example \ref{ex:subG} considers a $\sigma^2$-subGaussian distribution with known $\sigma^2$. Example \ref{ex:moment} investigates a distribution with a bounded $p$-th moment. While these examples construct the confidence sequence for the mean, Example \ref{ex:quantile} focuses on a quantile. Example \ref{ex:regression} considers regression coefficients in a linear model. Throughout the examples, we fix $i\in[k]$. In Examples \ref{ex:subG}--\ref{ex:quantile}, we suppose that data $X_{i1},X_{i2},\dots,$ are independent and identically distributed according to distribution $ P_i$. Define the filtration $\cF_{iT}=\sigma(X_{it}:t\leq T)$. 

\begin{ex}\normalfont\label{ex:subG}
Suppose that $P_i$ is a $\sigma^2$-subGaussian distribution with $\E[\exp\{\lambda(X_{i1}-\theta_i)\}]\leq\exp\{\sigma^2\lambda^2/2\}$ for all $\lambda\in\bbR$, where $\theta_i$ is the unknown mean and $\sigma^2$ is the known variance proxy. For any predictable sequence $(\lambda_{iT})_T$, both $M_{iT}^{\pm}=\prod_{t=1}^{T}\exp\{\mp\lambda_{it}(X_{it}-\theta_i)-\sigma^2\lambda_{it}^2/2\}$ are nonnegative supermartingales with respect to $\cF_{iT}$ with the expectations less than unity. Therefore, $M_{iT}^+$ and $M_{iT}^-$ satisfy Ville's inequality.
Inverting them with respect to $\theta_i$ yields $U_{iT}(\alpha)=V_{iT}+W_{iT}(\alpha)$ and $L_{iT}(\alpha)=V_{iT}-W_{iT}(\alpha)$, where $V_{iT}
=\sum_{t=1}^{T}\lambda_{it}X_{it}/\sum_{t=1}^{T}\lambda_{it}$ and $W_{iT}(\alpha)
=(\sigma^2\sum_{t=1}^{T}\lambda_{it}^2/2-\log \alpha)/\sum_{t=1}^{T}\lambda_{it}$. Then they satisfy \eqref{eq:bounded}. A discussion of the choice of $(\lambda_{iT})_T$ is given in Section \ref{sec:discussion_lambda} \citep{waudby2024estimating}. 
\end{ex}

\begin{ex}\normalfont\label{ex:moment}
Suppose that $P_i$ has a mean $\theta_i$ and a bounded absolute $p$-th central moment $\E[|X_{i1}-\theta_i|^p]\leq v$ for some $p\in(1,2]$ and $v>0$. Unlike Example \ref{ex:subG}, $P_i$ can be a heavy-tailed distribution with infinite higher-moments. For any predictable sequence $(\lambda_{iT})_T$, both
$M_{iT}^\pm=\prod_{t=1}^{T}\exp\{\mp\phi_p(\lambda_{it}(X_{it}-\theta_i))-\lambda_{it}^pv/p\}$, where $ \phi_p(x)=\sgn(x)\log(1+|x|+|x|^p/p)$, are nonnegative supermartingales with respect to $\cF_{iT}$ with the expectations less than unity \citep[Section 9]{wang2023catoni}. 
\end{ex}

\begin{ex}\normalfont\label{ex:quantile}
Suppose that $P_i$ is any marginal distribution with the $q$-th quantile $\theta_i$, which is of our interest. Define $\what{\theta}_{iT}^+(q)=\sup\{x:T^{-1}\sum_{t=1}^{T}\mathbb{I}\{X_{it}\leq x\}\leq q\}$ and $\what{\theta}_{iT}^-(q)=\sup\{x:T^{-1}\sum_{t=1}^{T}\mathbb{I}\{X_{it}\leq x\}<q\}$ be upper and lower empirical quantile functions, respectively. Then, 
$U_{iT}(\alpha)=\what{\theta}_{iT}^-(q+f_T(q))$ and 
$L_{iT}(\alpha)=\what{\theta}_{iT}^+(q-f_T(1-q))$, 
where $f_t(q)=1.5\sqrt{q(1-q)l(t)}+0.8l(t)$ with $l(t)=\{1.4\log\log(2.1t)+\log(5/\alpha)\}/t$, 
satisfy \eqref{eq:bounded}, regardless of the underlying distribution \citep{howard2022sequential}. 
\end{ex}

\begin{ex}\normalfont\label{ex:regression}
Consider a linear model $Y_T=X_T\beta+Z_T\theta_i+\epsilon_T$, where $Y_T\in\bbR^T$ is the vector of observations, $X_T\in\bbR^{T\times p}$ and $Z_T\in\bbR^T$ are design matrices, and $\epsilon_T\sim N(0,\sigma^2I_T)$. \cite{lindon2022anytime} constructed $U_{iT}(\alpha)$ and $L_{iT}(\alpha)$ for $\theta_i$ satisfying \eqref{eq:bounded} in a somewhat intricate manner; see Corollary 3.8 of their paper for the explicit construction.

\end{ex}

\section{Methodology}\label{sec:proc}
As our main contribution, we propose two statistical methodologies that will achieve the two goals presented in Section \ref{sec:problem}: SCS and PSI in Sections \ref{sec:proc_sequential} and \ref{sec:proc_post}, respectively. Throughout this and next sections, we suppose that, given data up to time $T\in\bbN$, two $(\cF_T)_{T\in\bbN}$-adapted sequences of statistics, $(U_{iT}(\alpha))_{T\in\bbN}$ and $(L_{iT}(\alpha))_{T\in\bbN}$, satisfying \eqref{eq:bounded} are available for the target parameter $\theta_i$ for each $i\in[k]$ and any $\alpha\in(0,1)$.

\subsection{Sequential correct screening}\label{sec:proc_sequential}

Our SCS is implemented as follows. 
\begin{proc}[SCS]\normalfont\label{proc1}
Set $\alpha_{km}=\alpha/\{2m(k-m)\}$ for any $\alpha\in(0,1)$. With initialization $\hat\cS_0=[k]$, construct
\begin{align}
\hat\cS_T=\hat\cS_{T-1}\backslash \left\{i: U_{iT}(\alpha_{km})<L_T^{(m)}(\alpha_{km})\right\}~~\text{for}~~T\in\bbN,
\end{align}
where $L_T^{(m)}(\alpha_{km})$ is the $m$-th largest order statistic among $\{L_{iT}(\alpha_{km}):i\in\hat\cS_{T-1}\}$.
\end{proc}

\begin{figure}[htb]
  \begin{minipage}[b]{0.5\linewidth}
    \centering
    \includegraphics[width=1\linewidth]{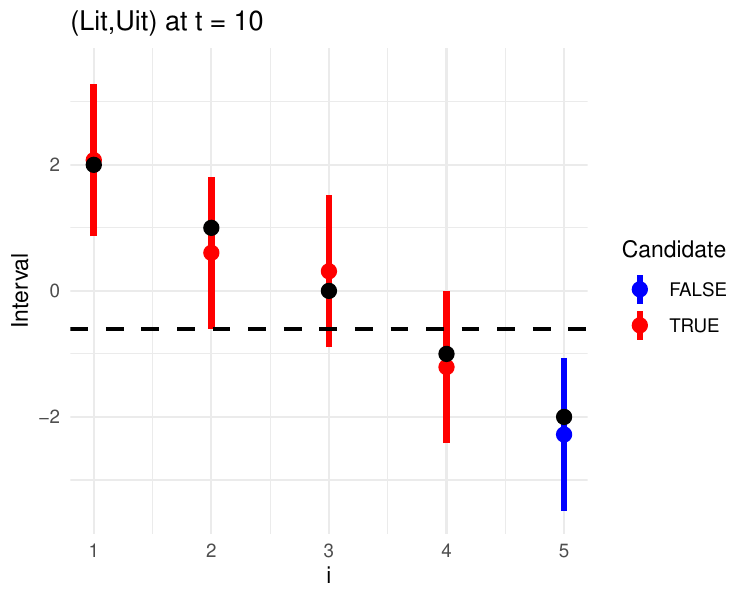}
  \end{minipage}
  \begin{minipage}[b]{0.5\linewidth}
    \centering
    \includegraphics[width=1\linewidth]{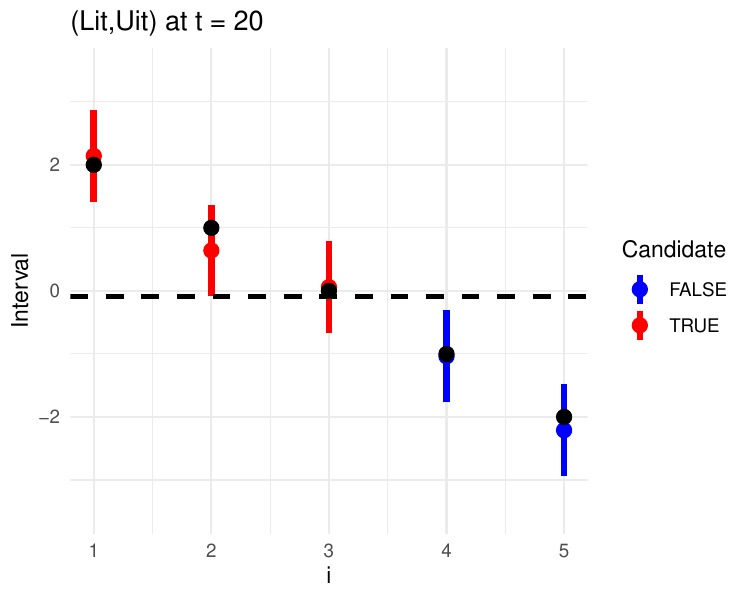}
  \end{minipage}
\caption{Plots of intervals $(L_{iT}(\alpha_{km}),U_{iT}(\alpha_{km}))$ for $i\in[k]$ at $T=10$ (left panel) and $T=20$ (right panel). Procedure \ref{proc1} is applied to find the $m=2$ largest means among $k=5$ normal distributions $N(\theta_i,1)$, where $\theta_1=2$, $\theta_2=1$, $\theta_3=0$, $\theta_4=-1$, and $\theta_5=-2$ (black dots). For each $i\in[k]$, we obtain data $X_{i1},X_{i2},\dots\sim N(\theta_i,1)$ independently. According to Example \ref{ex:subG} with $\lambda_{iT}=1/2$, we then calculate $U_{iT}(\alpha_{km})=\bar{X}_{iT}+h(T,\alpha_{km})$ and $L_{iT}(\alpha_{km})=\bar{X}_{iT}-h(T,\alpha_{km})$, where $\bar{X}_{iT}=(1/T)\sum_{t=1}^{T}X_{it}$ (red/blue dots) and $h(T,\alpha_{km})=1/4-(2/T)\log \alpha_{km}$. 
This construction satisfies \eqref{eq:bounded}. 
The threshold is given by $L_T^{(m)}(\alpha_{km})$ (black dashed lines); we set $i\not\in\hat{\cS}_T$ if the corresponding interval entirely lies below the line (blue), and $i\in\hat{\cS}_T$ otherwise (red). 
In consequence, we obtain $\hat{\cS}_{10}=\{1,2,3,4\}$ and $\hat{\cS}_{20}=\{1,2,3\}$.}
  \label{fig:demo}
\end{figure}

Procedure \ref{proc1} sequentially constructs shrinking subsets $\hat{\cS}_T\subset[k]$ to include $m$-promising $\cS$. With a carefully chosen level $\alpha_{km}$, it sequentially monitors the intervals $(L_{iT}(\alpha_{km}),U_{iT}(\alpha_{km}))$ and screen out $i$ from $\hat{\cS}_{T-1}$ if there is sufficient evidence that $\theta_i$ is small. More specifically, for each $i\in\hat{\cS}_{T-1}$, if $(L_{iT}(\alpha_{km}),U_{iT}(\alpha_{km}))$ entirely lies below $L_T^{(m)}(\alpha_{km})$, then we let $i\notin\hat{\cS}_T$. 
Notice that to avoid the event $\{|\hat{\cS}_T|<m\}$ occurs, $L_T^{(m)}(\alpha_{km})$ should be the $m$-th largest in $\{L_{iT}(\alpha_{km}):i\in\hat{\cS}_{T-1}\}$ but not in $\{L_{iT}(\alpha_{km}):i\in[k]\}$. Figure \ref{fig:demo} illustrates Procedure \ref{proc1} applied to a simple situation. 

Once screened out, the $i$-th variable will never be a member of $\hat{\cS}_T$ in future. By the construction, $\hat{\cS}_T$ will not change once the number of candidates becomes $|\hat{\cS}_T|=m$. Moreover, even when $|\hat{\cS}_T|>m$, we can stop updating the screened set whenever we want, and set the last $T$ as $\tau$. This property is called the anytime validity (sequential property), and is due to \eqref{eq:bounded}. A formal theory is given in Section \ref{sec:theory_correct}.

\subsubsection{Comparison to the LUCB algorithm}\label{sec:LUCB}

In the literature of BAI, a similar methodology, called the LUCB algorithm, has been proposed \citep{kalyanakrishnan2012pac}. 
Like our Procedure \ref{proc1}, it requires some well-constructed intervals $(L_{iT}^{\textsf{LUCB}},U_{iT}^{\textsf{LUCB}})$ for $\theta_i$, though they are different from \eqref{eq:bounded}, to seek the $m$-promising parameters; see Section 3 of their paper for more details. 

Suppose such intervals are given. Letting $\hat{\theta}_{iT}$ and $\hat{\theta}_{(m)T}$ denote an estimator of $\theta_i$ at time $T$ and the $m$-th largest among them, respectively, we define $\mathcal{L}_T=[k]\backslash\mathcal{U}_T$ with $\mathcal{U}_T=\{i\in[k]:\hat{\theta}_{iT}\geq\hat{\theta}_{(m)T}\}$. Then the LUCB stops at time $\tau$, where $\tau$ is a stopping time given by
\begin{align}
    \tau=\min\left\{T\in\bbN:\max_{i\in\mathcal{L}_T}U_{iT}^{\textsf{LUCB}}<\min_{i\in\mathcal{U}_T}L_{iT}^{\textsf{LUCB}}\right\}, \label{LUCB_stop}
\end{align}
and declares that $\hat{\cS}^{\textsf{LUCB}}=\mathcal{U}_\tau$ is the set of $m$-promising parameters.

Despite the similarity of LUCB to Procedure \ref{proc1}, it does not achieve Property \ref{property}. The primary distinction between LUCB and our approach lies in the strategy for identifying the $m$-promising parameters; the LUCB performs a one-shot selection based on fully accumulated information, while our method adopts sequential screening to eliminate parameters that are unlikely to be $m$-promising according to the information available at each time point. Consequently, LUCB requires continued sampling until a stopping criterion in \eqref{LUCB_stop} is met, at which point it returns its output, $\hat{\cS}^{\textsf{LUCB}}$. In contrast, our method outputs a reasonably sized candidate set $\hat{\cS}_T$ at each time $T$, based on the information currently available. 
While such differences exist, we will compare the performances of Procedure \ref{proc1} to a slightly modified LUCB (without theoretical guarantee) due to its similarity to our method in Section \ref{sec:simulations}; see also Section \ref{sec:related}.

\subsection{Post-screening inference} \label{sec:proc_post}
We propose our PSI under the same situation as in Procedure \ref{proc1}.
As discussed in Section \ref{sec:post-selection}, we conduct PSI at any $(\cF_T)_{T\in\bbN}$-adapted stopping time $\tau$.

\begin{proc}[PSI]\normalfont\label{proc2}
Given $\hat{\cS}_\tau$ by Procedure \ref{proc1}, set $\tilde{\alpha}=\alpha|\hat{\cS}_\tau|/(2k)$. Construct the adjusted CI as $(L_{i\tau}(\tilde{\alpha}),U_{i\tau}(\tilde{\alpha}))$  for each $\{\theta_i:i\in\hat{\cS}_\tau\}$.
\end{proc}

\begin{proc}[Bonferroni-PSI]\normalfont\label{proc3}
Given $\hat{\cS}_\tau$ by Procedure \ref{proc1}, set $\alpha^\textsf{B}=m\alpha/(2k)$. Construct the adjusted CI as $(L_{i\tau}(\alpha^\textsf{B}),U_{i\tau}(\alpha^\textsf{B}))$  for each $\{\theta_i:i\in\hat{\cS}_\tau\}$.
\end{proc}

Procedures \ref{proc2} and \ref{proc3} constructs the adjusted CI for each screened parameters, $\{\theta_i:i\in\hat{\cS}_\tau\}$, with controlling the sFCR after conducting the SCS. The intervals obtained by Procedure \ref{proc1}, $(L_{i\tau}(\alpha_{km}),U_{i\tau}(\alpha_{km}))$, do not necessarily control the sFCR. As a remedy, Procedures \ref{proc2} and \ref{proc3} aims to construct intervals that always control the sFCR. Since $\tilde{\alpha}\leq\alpha^\textsf{B}$, Procedure \ref{proc2} constructs narrower intervals than Procedure \ref{proc3}. However, to control sFCR, Procedure \ref{proc2} requires additional assumption, while Procedure \ref{proc3} does not. A formal theory is given in Section \ref{sec:theory_post}.

\subsubsection{Comparison to the e-BY procedure}\label{sec:e-BY}

We briefly review the method, called the e-BY for sFCR control \citep{xu2024post}. This forms a basis of theoretical justification for Procedure \ref{proc2}. 
To begin with, we introduce an \textit{e-process}, defined as 
a process of nonnegative random variables $(E_T)_{T\in\bbN}$ such that $\E[E_\tau]\leq1$ for any stopping time $\tau$ with respect to filtration $(\cF_T)_{T\in\bbN}$.

For each $i\in[k]$, suppose that we have a family of e-processes $\{(E_{iT}(\theta_i))_{T}:\theta_i\in\bbR\}$, in a sense that $E_{iT}(\theta_i)$ is an e-process for any stopping time $\tau$ in a sense that $\E[E_{iT}(\theta_i)]\leq1$ for any stopping time $\tau$ when the expectation is taken under $\theta_i$. Then, the stopped interval
\begin{align}
    C_{i\tau}(\alpha)=\left\{\theta_i\in\bbR:E_{i\tau}(\theta_i) < 1/\alpha \right\}
\end{align}
is called the e-CI. Indeed, this is also a CI because of Markov's inequality and the definition of e-processes.
If we have access to e-CIs for each parameter and to any stopping time $\tau$, then we can conduct the e-BY as follows. First we select a subset of parameters, $\hat{\cS}_\tau\subset\{\theta_1,\dots,\theta_k\}$, by any method. Then construct $C_{i\tau}(\tilde{\alpha}^{\text{e-BY}})$ with $\tilde{\alpha}^{\text{e-BY}}=\alpha|\hat{\cS}_\tau|/k$ for each selected parameter in $\hat{\cS}_\tau$. \cite{xu2024post} showed that the sFCR of $C_{i\tau}(\tilde{\alpha}^{\text{e-BY}})$ given $\hat{\cS}_\tau$ is controlled at most $\alpha$, 
regardless of the selection rule and the dependence structure among $C_{i\tau}(\tilde{\alpha}^{\text{e-BY}})$s.

Regarding the difference, while Procedure \ref{proc2} relies on \eqref{eq:bounded}, the e-BY is based on e-processes. This construction is essential for sFCR control via the e-BY. In Section \ref{sec:theory_post}, we develop a theoretical guarantee for sFCR control of Procedure \ref{proc2} by linking it to the e-BY framework.

\section{Statistical Theory}\label{sec:theory}
In Section \ref{sec:theory_correct}, we show that Procedure \ref{proc1} achieves Property \ref{property} under some mild conditions. To make the asymptotics in Property \ref{property}(iii) hold, $U_{iT}(\alpha)$ and $L_{iT}(\alpha)$ have to be carefully constructed. We provide a discussion of this issue in Section \ref{sec:discussion_lambda}. Section \ref{sec:theory_post} shows that Procedures \ref{proc2} and \ref{proc3} control the sFCR. Finally, in Section \ref{sec:interval}, we compare three intervals constructed by Procedures \ref{proc1}--\ref{proc3} in terms of the length.

Throughout this section, 
we suppose that $(\cF_T)_{T\in\bbN}$-adapted sequences of statistics, $(U_{iT}(\alpha))_{T\in\bbN}$ and $(L_{iT}(\alpha))_{T\in\bbN}$, satisfying \eqref{eq:bounded} for each $i\in[k]$ and any $\alpha\in(0,1)$ are available.

\subsection{Theory for sequential correct screening}\label{sec:theory_correct}
We show that our SCS in Procedure \ref{proc1} satisfies Property \ref{property}(i)--(iii). 

\begin{con}\normalfont\label{con:promising}
There are exactly $m$ $m$-promising parameters, ordered as $\theta_1\geq\dots\geq\theta_m>\theta_{m+1}\geq\dots\geq\theta_k$.
\end{con}

\begin{con}\normalfont\label{con:lln}
For any $i\in[k]$ and $\alpha\in(0,1)$, we have $U_{iT}(\alpha)\xrightarrow{p}\theta_i$ and $L_{iT}(\alpha)\xrightarrow{p}\theta_i$ as $T\to\infty$.
\end{con}
Condition \ref{con:promising} stipulates the members of the set of $m$-promising parameters. This assumes the existence of an IZ between $\theta_m$ and $\theta_{m+1}$, but its magnitude is not required to be known. Under this condition, we have $\cS=[m]$. 
Condition \ref{con:lln} says that for each $i\in[k]$ the interval $(L_{iT}(\alpha),U_{iT}(\alpha))$ degenerates in probability to the true parameter $\theta_i$ pointwise in $\alpha$. This condition holds in many cases; a related discussion is given in Section \ref{sec:discussion_lambda}.

We first verify that SCS in Procedure \ref{proc1} achieves Property \ref{property} (i)--(iii).

\begin{thm}\label{thm:correct}Suppose that the sequence $(\hat{\cS}_T)_{T\in\bbN}$ is constructed by SCS in Procedure \ref{proc1}. Then the following statements are true:
\begin{enumerate}
\item[(a)] Property \ref{property}(i) holds. 
\item[(b)] If Condition \ref{con:promising} is satisfied, Property \ref{property}(ii) holds.
\item[(c)] If Conditions \ref{con:promising} and \ref{con:lln} are satisfied, Property \ref{property}(iii) holds.
\end{enumerate}
\end{thm}
The proof of Theorem \ref{thm:correct} is provided in Supplementary Material. The theorem holds under the very mild conditions. Specifically, any dependence structure in $U_{iT}(\alpha)$ and $L_{iT}(\alpha)$ is allowed.

\subsubsection{Discussion of Condition \ref{con:lln}}\label{sec:discussion_lambda}

We discuss a sufficient condition for Condition \ref{con:lln} under a sub-Gaussian (subG) assumption. For any $\alpha\in(0,1)$, the statistics $U_{iT}(\alpha)$ and $L_{iT}(\alpha)$ in Example \ref{ex:subG} satisfy Condition \ref{con:lln} if 
$V_{iT}\to_p \theta_i$ and $W_{iT}(\alpha)\to_p 0$ with any predictable sequence $(\lambda_{iT})_T$  with respect to $(\cF_T)_T$. To see this, set $\lambda_{iT}\asymp_p 1/\sqrt{T\log T}$, which yields $\sum_{t=1}^T\lambda_{it}\asymp_p \sqrt{T/\log T}$ and  $\sum_{t=1}^T\lambda_{it}^2/\sum_{t=1}^T\lambda_{it} \to_p 0$. Thus the second convergence clearly holds. The first convergence is also verified by the law of large numbers and Theorem 1 of \cite{etemadi2006convergence}. Although this choice of $\lambda_{iT}$ leads to a good shrinkage rate, it may not be preferable in many applications because of the small sample sizes \citep[Section 3.5]{howard2021time}. An extension to the other cases may be achieved in a similar manner. We refer to \cite{waudby2024estimating} for a related discussion.

\subsection{Theory for post-screening inference}\label{sec:theory_post}
We verify that the adjusted CIs by Procedures \ref{proc2} and \ref{proc3} control the sFCR, denoted respectively by $\text{sFCR}(\tilde{\alpha})$ and $\text{sFCR}(\alpha^\textsf{B})$ with
\begin{align}
\text{sFCR}(\alpha)=\E\left[\sum_{i\in\hat{\cS}_\tau}\frac{\mathbb{I}\{\theta_i\notin (L_{i\tau}(\alpha),U_{i\tau}(\alpha))\}}{|\hat{\cS}_\tau|}\right],
\end{align}
to be at most $\alpha$. Section \ref{sec:interval} compares the lengths of such intervals obtained by Procedures \ref{proc1}--\ref{proc3}. Specifically, it clarifies when the intervals of Procedure \ref{proc2} become narrower. 

First we provide a sufficient condition to guarantee sFCR control of Procedure \ref{proc2}.

\begin{con}\normalfont\label{con:posthoce}
    For any $i\in[k]$, $\alpha\in(0,1)$ and any stopping time $\tau$ with respect to $(\cF_T)_T$, there exist $E_{i\tau}^+$ and $E_{i\tau}^-$ such that a.s.\ $\E[E_{i\tau}^+]\leq1$, $\E[E_{i\tau}^-]\leq1$,
    \begin{align}
        \mathbb{I}\{\theta_i\leq L_{i\tau}(\alpha)\}\leq\mathbb{I}\left\{E_{i\tau}^-\geq1/\alpha\right\},~\text{and}~\mathbb{I}\{\theta_i\geq U_{i\tau}(\alpha)\}\leq\mathbb{I}\left\{E_{i\tau}^+\geq1/\alpha\right\}.
    \end{align}
\end{con}
Condition \ref{con:posthoce} implies that the interval $(L_{i\tau}(\alpha),U_{i\tau}(\alpha))$ includes some e-CI. This condition holds in many cases. For example, in Example \ref{ex:subG}, Condition \ref{con:posthoce} holds with equality by letting $E_{iT}^-=M_{iT}^-$ and $E_{iT}^+=M_{iT}^+$. However, even when we have $(U_{iT}(\alpha))_T$ and $(L_{iT}(\alpha))_T$ satisfying \eqref{eq:bounded}, Condition \ref{con:posthoce} does not necessarily hold. Indeed, there exist $(U_{iT}(\alpha))_T$ and $(L_{iT}(\alpha))_T$ that are based on extended nonnegative supermartingales whose expectations can be infinite; see \cite{wang2024anytime, wang2025extended}.

\begin{thm}\label{thm:FCR}
Suppose that Condition \ref{con:posthoce} holds. Then for any $\alpha\in(0,1)$ and any stopping time $\tau$ with respect to $(\cF_T)_{T\in\bbN}$, we have $\text{sFCR}(\tilde{\alpha})\leq \alpha$.
\end{thm}

\begin{thm}\label{thm:FCR1}
For any $\alpha\in(0,1)$ and any stopping time $\tau$ with respect to $(\cF_T)_{T\in\bbN}$, we have $\text{sFCR}(\alpha^\textsf{B})\leq \alpha$.
\end{thm}
The proofs are collected in Supplementary Material. In contrast to Theorem \ref{thm:FCR} for Procedure \ref{proc2}, Theorem \ref{thm:FCR1} says that Procedure \ref{proc3} controls sFCR without requiring Condition \ref{con:posthoce}.

\subsubsection{Discussion of the lengths of intervals}\label{sec:interval}
Given $\alpha$, we compare the lengths of three intervals, $(L_{iT}(a),U_{iT}(a))$ for $a\in \{\alpha_{km},\tilde{\alpha},\alpha^\textsf{B}\}$. They are constructed from Procedure \ref{proc1}--\ref{proc3}, respectively. In the comparison, we suppose that $(L_{iT}(a),U_{iT}(a))\subset(L_{iT}(b),U_{iT}(b))$ for any $1>a\geq b>0$. This makes the discussion simpler since the comparison of the lengths reduces to the comparison of $\alpha_{km}$, $\tilde{\alpha}$ and $\alpha^{\textsf{B}}$. 

\begin{thm}\label{thm:interval}
All the following hold:
    \begin{enumerate}
        \item[(a)] If $m=1$ and $k\geq 2$, and
        \begin{enumerate}
            \item[i.] $|\hat{\cS}_t|=1$, then $\alpha^{\textsf{B}}=\tilde{\alpha}<\alpha_{km}$;
            \item[ii.] $|\hat{\cS}_t|\geq2$, then $\alpha^{\textsf{B}}<\alpha_{km}\leq\tilde{\alpha}$;
        \end{enumerate}
        \item[(b)] If $2\leq m\leq k-2$ and $k\geq 4$, then $\alpha_{km}\leq\alpha^{\textsf{B}}\leq\tilde{\alpha}$;
        \item[(c)] If $m=k-1$ and $k\geq 3$, then $\alpha_{km}<\alpha^{\textsf{B}}\leq\tilde{\alpha}$.
    \end{enumerate}
\end{thm}

The proof of Theorem \ref{thm:interval} is given in Supplementary Material. We see that $\tilde{\alpha}$ is the largest except 1a. In the case of 1a, however, whether the sFCR of $(L_{iT}(\alpha_{km}),U_{iT}(\alpha_{km}))$ is controlled or not is unclear.

\section{Simulation Studies}\label{sec:simulations}
For each $i\in[k]$, we generate data $\{X_{it}\}_{t=1}^T\sim\text{i.i.d.\ Ber}(\theta_i)$ for $T\in\{100,\ 1000,\ 10000\}$, where the $k$ parameters are specified by $\theta_i=1-i/k$. We set $k=50$, $m=3$, and $\alpha=0.1$ throughout this section. Since $X_{it}$ are bounded in $[0,1]$, Example \ref{ex:subG} suggests constructing
\begin{align}\label{eq:simulationUL}
        U_{iT}(\alpha),\ L_{iT}(\alpha)=\frac{\sum_{t=1}^{T}\lambda_{it}X_{it}}{\sum_{t=1}^{T}\lambda_{it}} \pm \frac{\sum_{t=1}^{T}(\lambda_{it}^2/8)-\log\alpha}{\sum_{t=1}^{T}\lambda_{it}},
\end{align}
where $\lambda_{it}=\sqrt{8\log(1/\alpha)/\{t\log(t+1)\}}\land1$. They satisfy \eqref{eq:bounded} and Condition \ref{con:lln}; see \cite{waudby2024estimating}.

\subsection{Simulation studies for SCS}

First, we demonstrate how well the screened sets $(\hat{\cS}_T)_T$ obtained by Procedure \ref{proc1} fulfill Property \ref{property}. 
To evaluate the performance, we would like to compare it with some existing methods. However, to the best of our knowledge, there is no procedure that is designed to satisfy Property \ref{property} as ours is. Therefore, we borrow a slightly modified version of the LUCB interval: 
\begin{align}
    \tilde{U}_{iT}(\alpha),\ \tilde{L}_{iT}(\alpha)
    &=\frac{1}{T}\sum_{t=1}^{T}X_{it}\pm\sqrt{\frac{1}{2T}\log\left(\frac{5k^5T^4}{4\alpha}\right)}. \label{eq:mod_LUCB}
\end{align}
We then run Procedure \ref{proc1} with \eqref{eq:mod_LUCB}. Thanks to the modification, the resulting set satisfies Property \ref{property}(i) with selecting $m$ or more parameters, but it is unclear whether this procedure fulfills Property \ref{property}(ii)--(iii). As mentioned in Section \ref{sec:LUCB}, LUCB is not originally intended for sequential screening, and hence the comparison may not be entirely fair. Still, we use the modified version as a benchmark just for illustrative purposes.

\begin{figure}[H]
    \centering
    \includegraphics[width=1\linewidth]{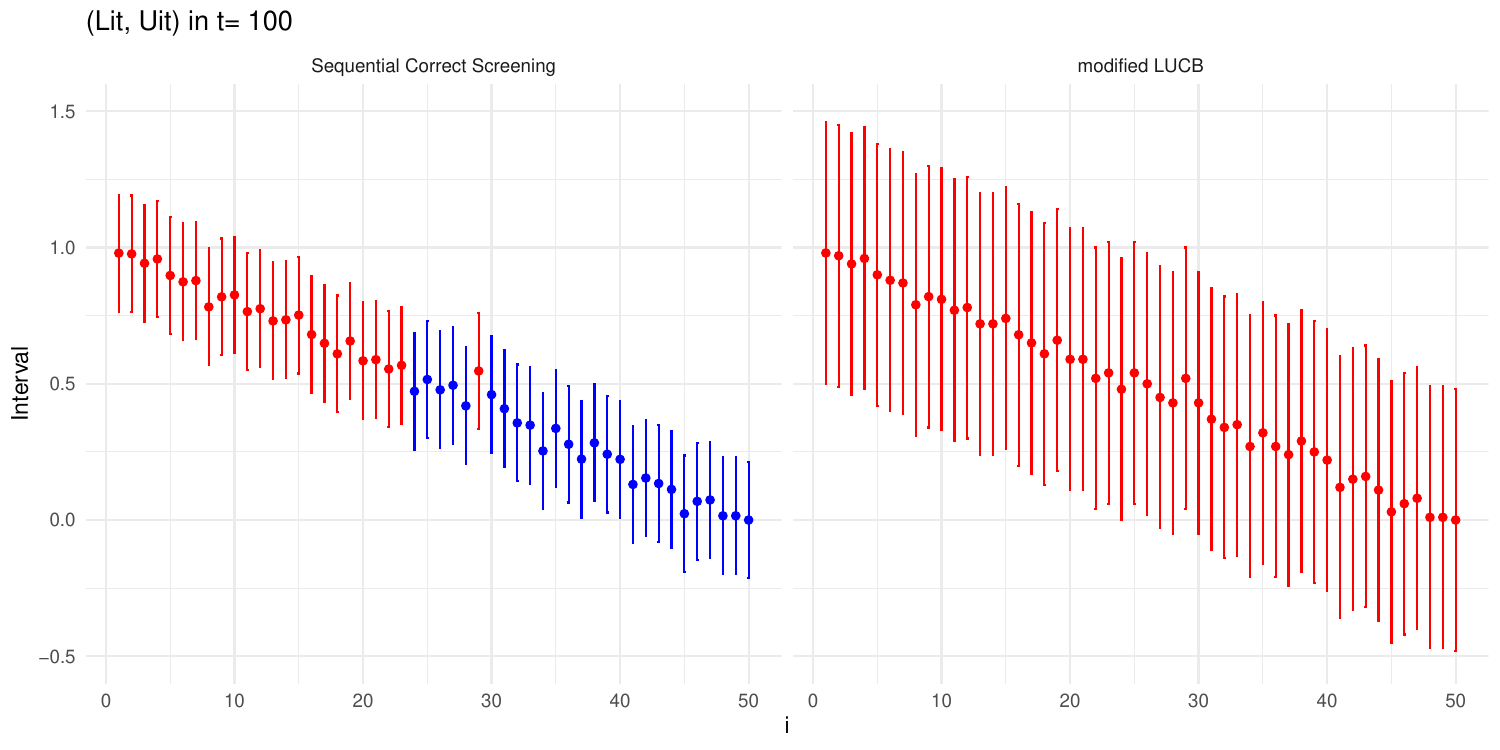}
    \caption{Comparison of the intervals constructed by Procedure \ref{proc1} (left) and the modified LUCB (right) with $T=100$.}
    \label{fig:simulation_t100}
\end{figure}

\begin{figure}[H]
    \centering
    \includegraphics[width=1\linewidth]{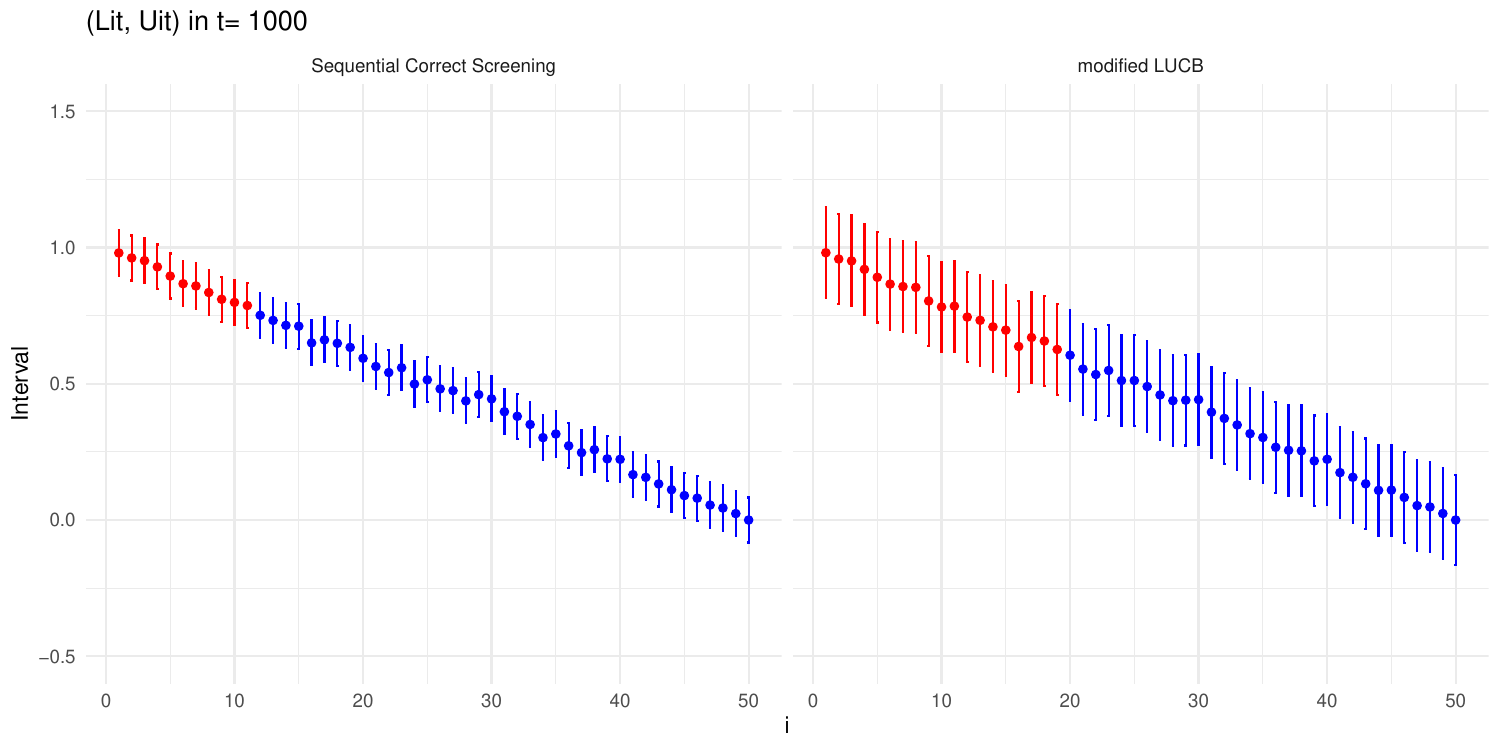}
    \caption{Comparison of the intervals constructed by Procedure \ref{proc1} (left) and the modified LUCB (right) with $T=\text{1,000}$.}
    \label{fig:simulation_t1000}
\end{figure}

\begin{figure}[H]
    \centering
    \includegraphics[width=1\linewidth]{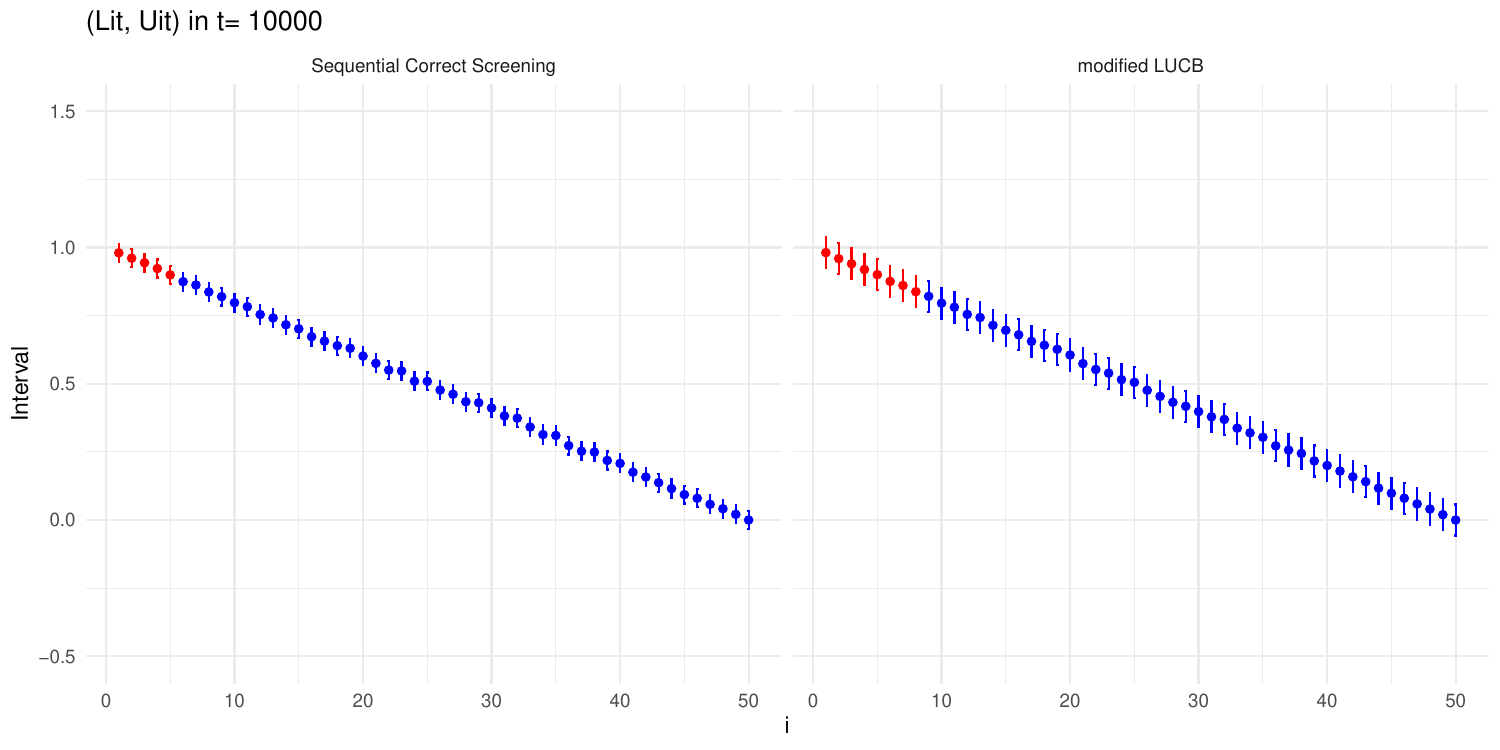}
    \caption{Comparison of the intervals constructed by Procedure \ref{proc1} (left) and the modified LUCB (right) with $T=\text{10,000}$.}
    \label{fig:simulation_t10000}
\end{figure}

\begin{figure}[H]
    \centering
    \includegraphics[width=1\linewidth]{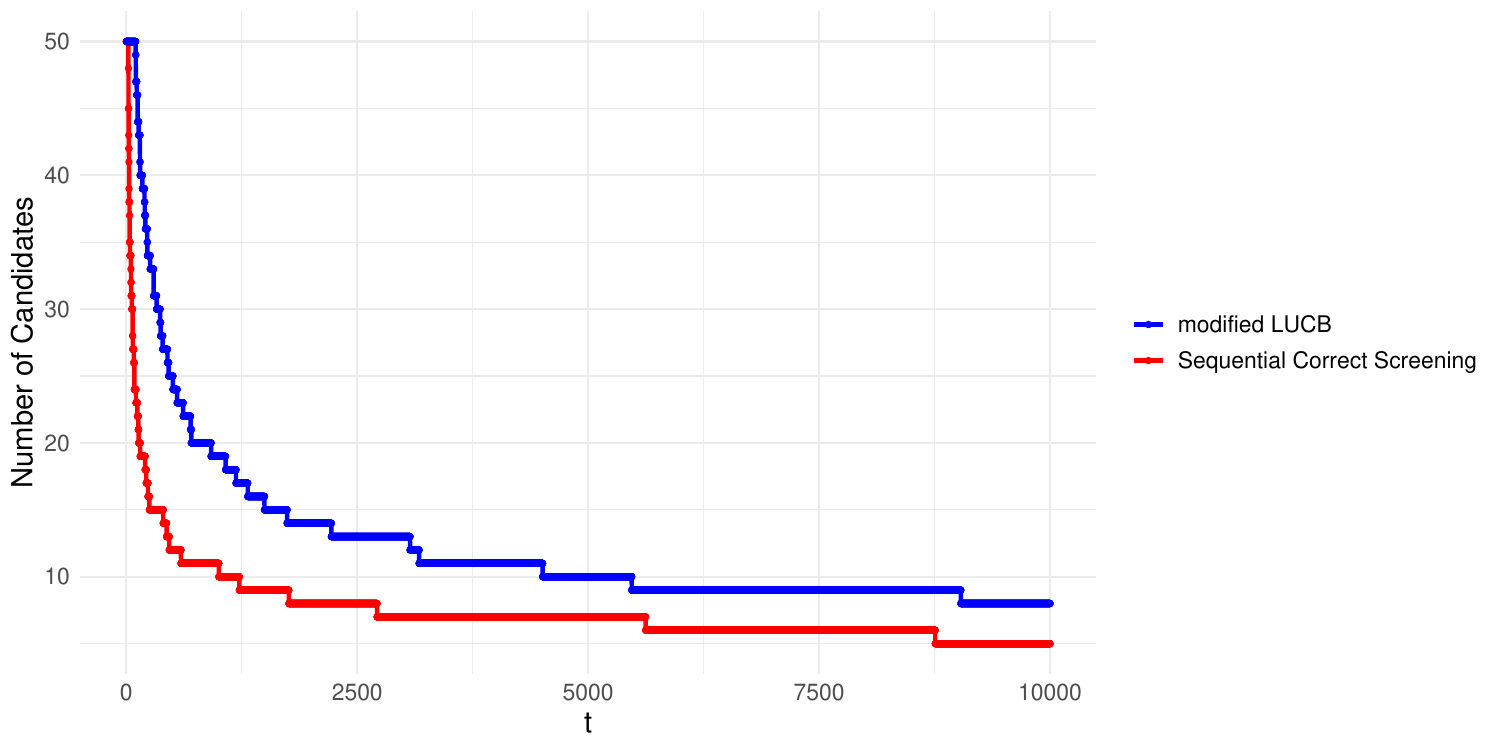}
    \caption{Changes in the number of screened candidates.}
    \label{fig:change}
\end{figure}

The comparison results are shown in Figures \ref{fig:simulation_t100}--\ref{fig:change}.
In Figures \ref{fig:simulation_t100}--\ref{fig:simulation_t10000}, the left and right panels indicate the intervals given by \eqref{eq:simulationUL} and \eqref{eq:mod_LUCB}, respectively, and the dots represent their first terms. At each time $t$, the intervals corresponding to the screened parameters are colored red, and blue otherwise. These figures show that Property \ref{property}(i)--(ii) hold for both procedures. 
Furthermore, Procedure \ref{proc1} is more powerful than the modified LUCB; both the number of screened parameters and the interval widths of ours shrink faster. Figure \ref{fig:change}, which depicts the red and blue curves representing the number of selected parameters of each procedure, supports this observation. Although omitted in the figure, we find that Property \ref{property}(iii) also holds for Procedure \ref{proc1}. 

\subsection{Simulation studies for PSI}

\begin{figure}[H]
    \centering
    \includegraphics[width=1\linewidth]{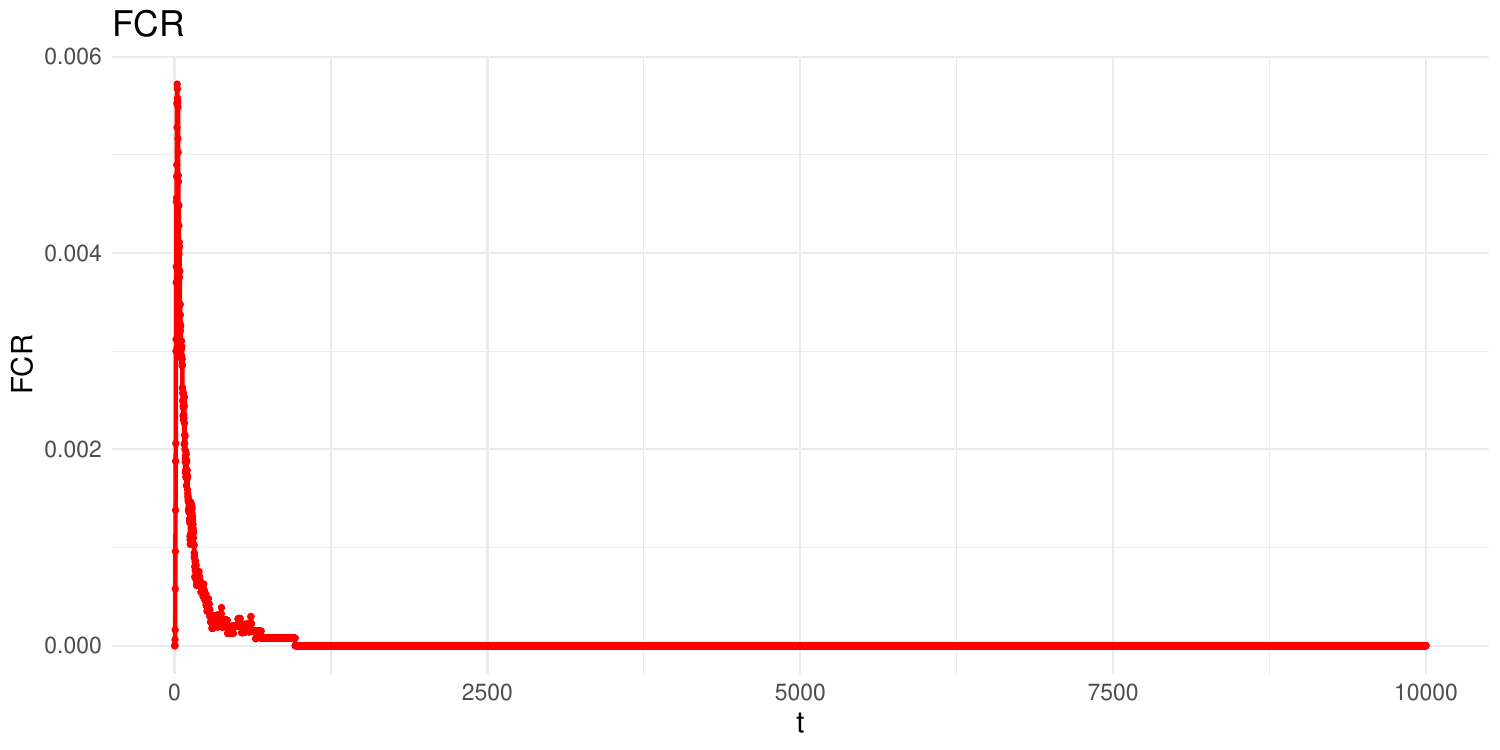}
    \caption{Plot of FCR of Procedure \ref{proc2}.}
    \label{fig:FCR_simulation}
\end{figure}

Finally, we check if the FCR is controlled by  Procedure \ref{proc2} with \eqref{eq:simulationUL}. For each $t$, we report the FCP:
\begin{align}
\text{FCP}_T=\sum_{i\in\hat{\cS}_T}\frac{\mathbb{I}\{\theta_i\notin (L_{iT}(\tilde{\alpha}),U_{iT}(\tilde{\alpha}))\}}{|\hat{\cS}_T|}.
\end{align}
We replicate $\text{1,000}$ times and take their average as the estimate of $\text{FCR}_T$ for each $T$. The result is shown in Figure \ref{fig:FCR_simulation}. It verifies that Procedure \ref{proc2} always controls FCR with target level $\alpha=0.1$ though it looks very conservative. 

\section{Empirical Application}\label{sec:applications}

We apply our SCS and PSI to the suicide rates dataset, obtained from Kaggle website\footnote{\url{https://www.kaggle.com/datasets/russellyates88/suicide-rates-overview-1985-to-2016/data}}, to explore which countries exhibit higher levels of suicide. The dataset $\{X_{it}\}$ contains the number of suicides of country $i\in[k]$ in year $t\in[T]$ with $k=101$ and $T=32$. For each $i\in[k]$, we suppose that $X_{it}$ is sampled from a conditional $\sigma^2$-subGaussian distribution with $\E[\exp\left\{\lambda(X_{it}-\theta_i)\right\}\mid X_{i1},\dots,X_{i,t-1}]\leq\exp\left\{\lambda^2\sigma^2/2\right\}$ a.s., where $\theta_i$ is the mean for each $i\in[k]$. We set $\alpha=0.1$, $\sigma=5$, $m=3$, and $\lambda=0.15$.

Some data points are missing in the dataset, but we use only observed data up to year $T$ to construct the statistics for the target parameters, $\theta_i$. Specifically, let $\bar{X}_{iT}=(1/T_i)\sum_{t=1}^{T}X_{it}\mathbb{I}\{\text{$X_{it}$ is observed}\}$ with $T_i=\sum_{t=1}^{T}\mathbb{I}\{\text{$X_{it}$ is observed}\}$ denote the sample mean of observed data. The list of all countries and their respective sample sizes can be found in Table \ref{table:suicide} of Section \ref{sec:table} in Supplementary Material.

\begin{figure}[H]
    \centering
    \includegraphics[width=1\linewidth]{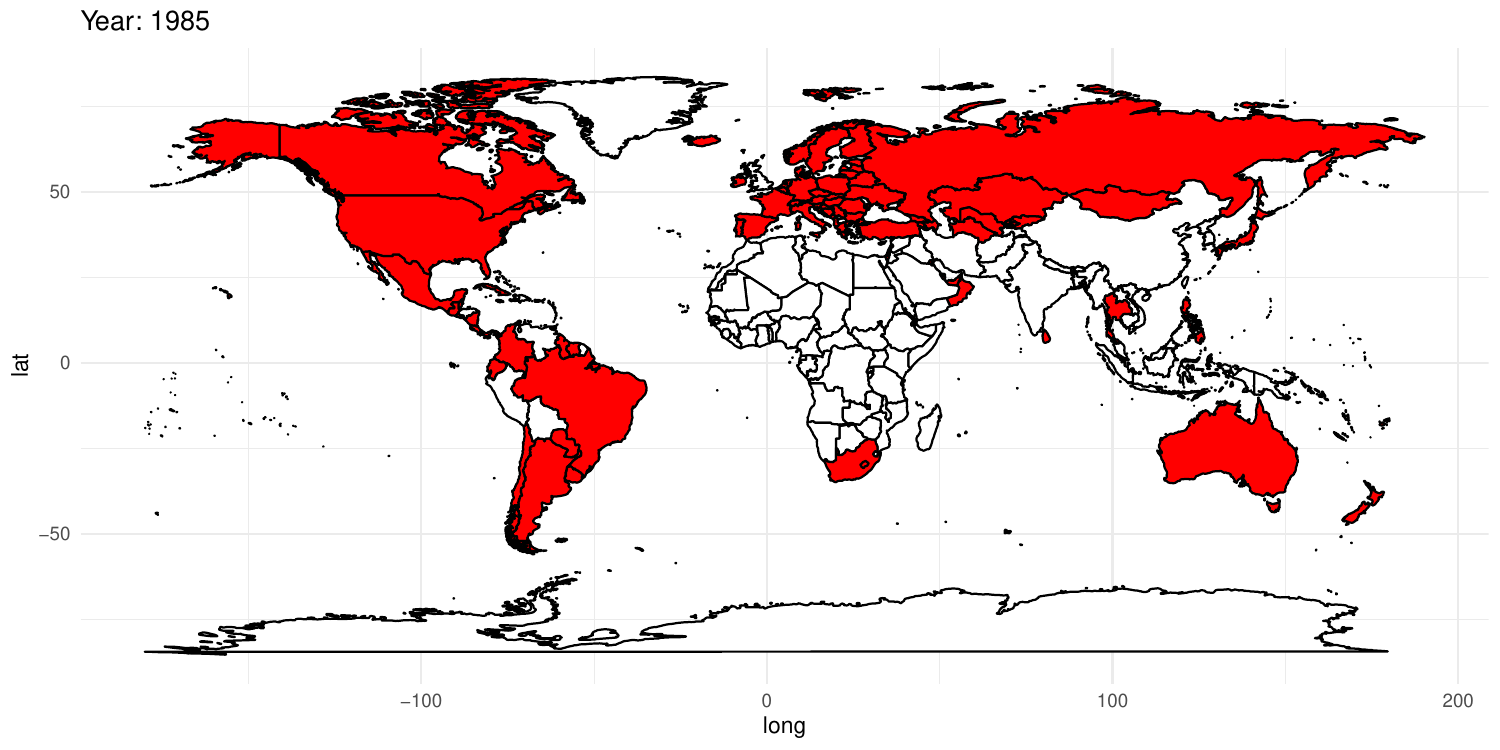}
    \caption{Selected countries in year 1985.}
    \label{fig:suicide_1985}
\end{figure}

\begin{figure}[H]
    \centering
    \includegraphics[width=1\linewidth]{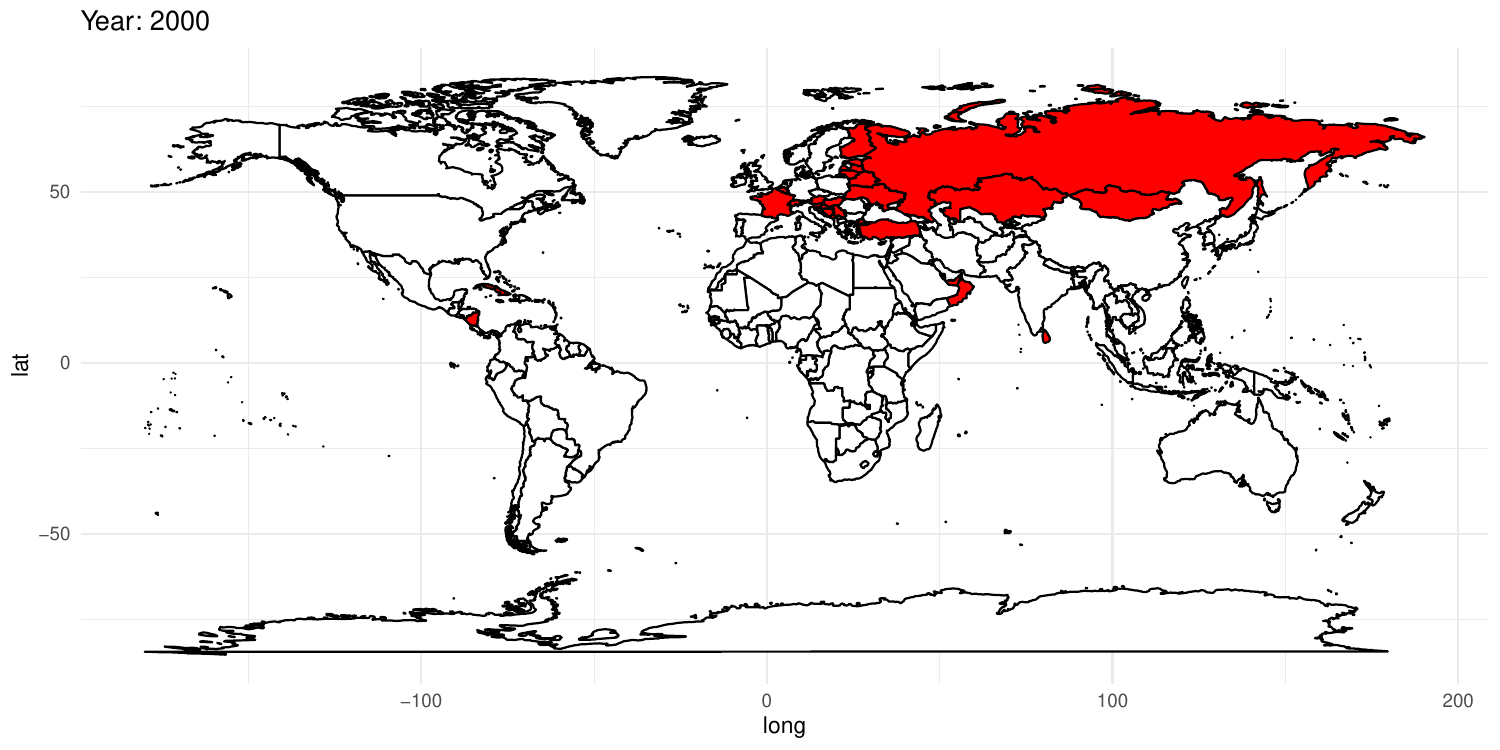}
    \caption{Selected countries in year 2000.}
    \label{fig:suicide_2000}
\end{figure}

\begin{figure}[H]
    \centering
    \includegraphics[width=1\linewidth]{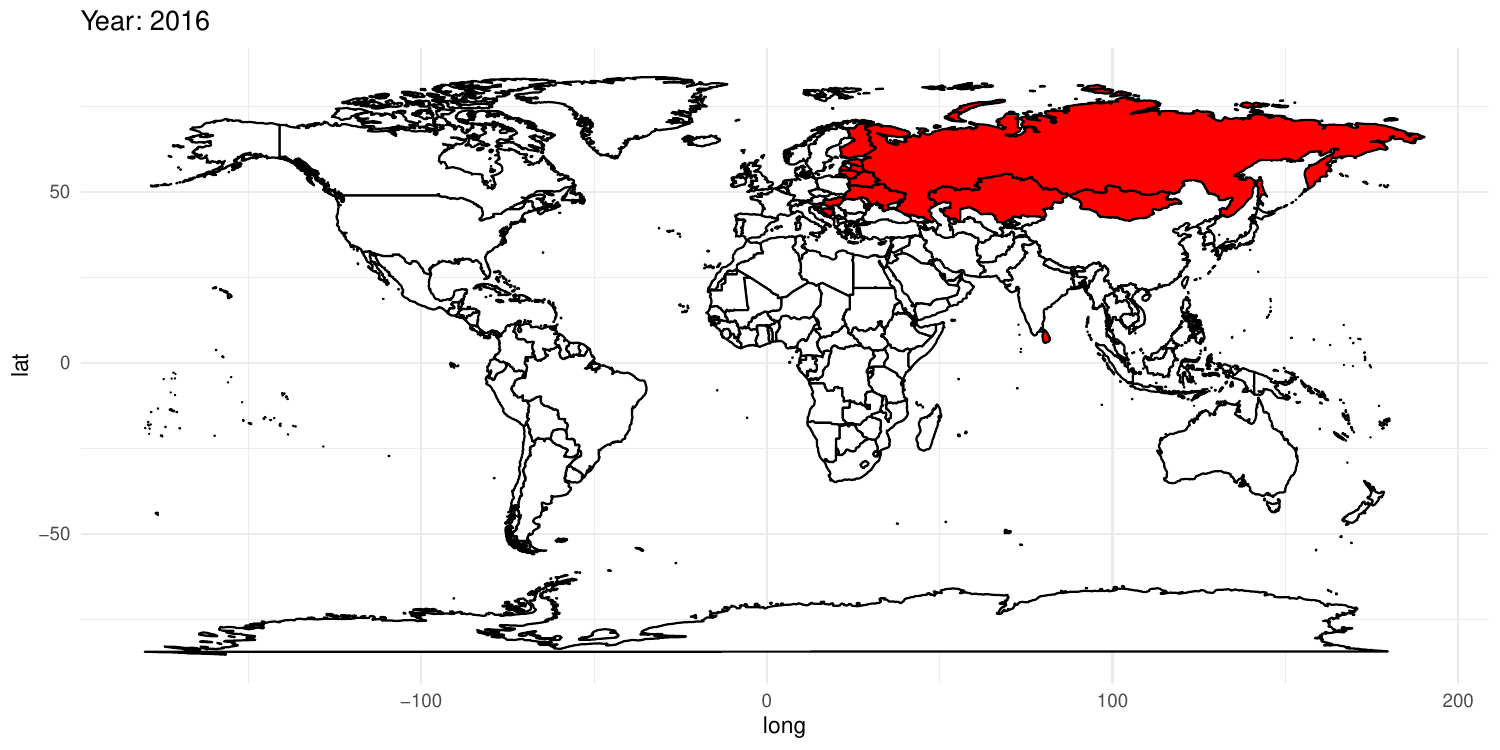}
    \caption{Selected countries in year 2016.}
    \label{fig:suicide_2016}
\end{figure}

\begin{figure}[H]
    \centering
    \includegraphics[width=1\linewidth]{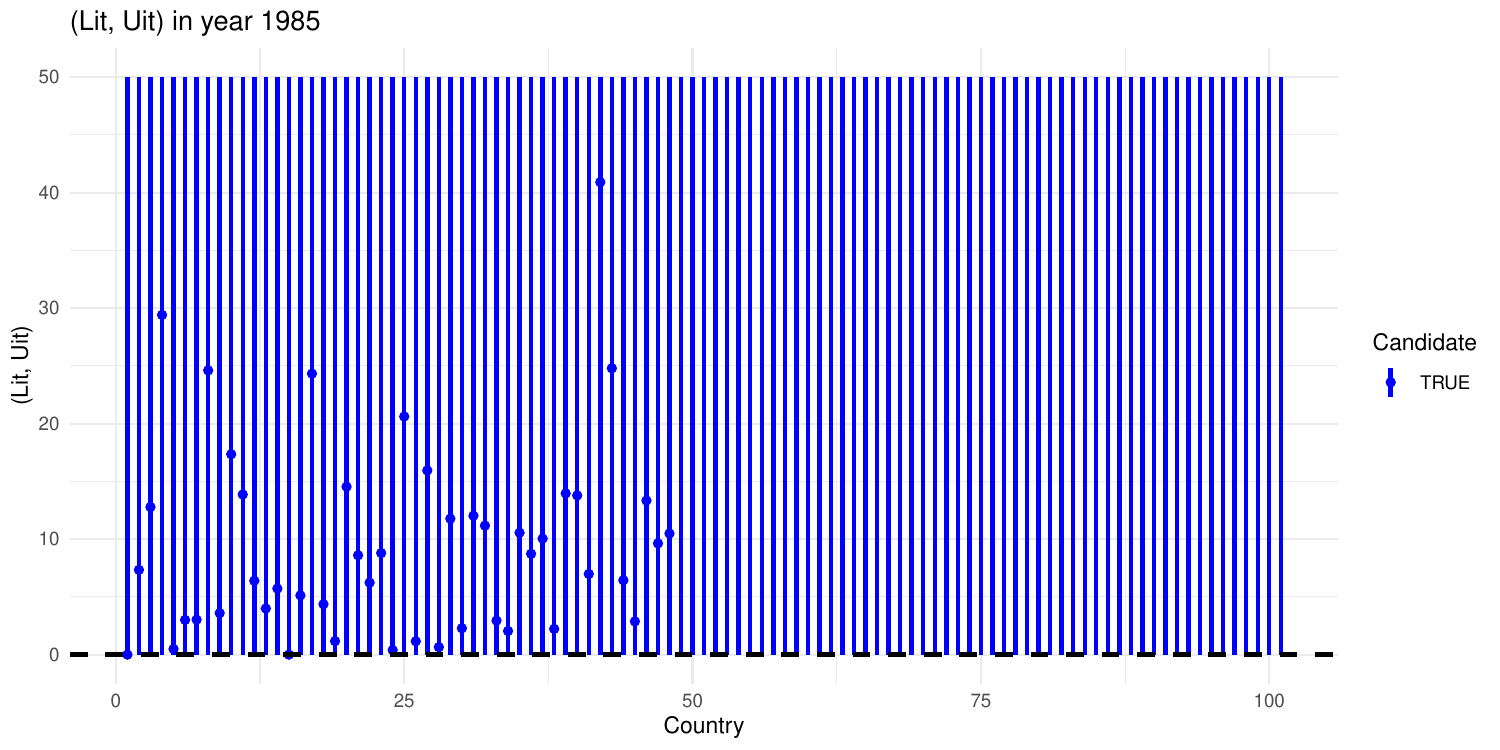}
    \caption{$(L_{it}(\alpha_{km}),U_{it}(\alpha_{km}))$ in year 1985.}
    \label{fig:suicide_ci_1985}
\end{figure}

\begin{figure}[H]
    \centering
    \includegraphics[width=1\linewidth]{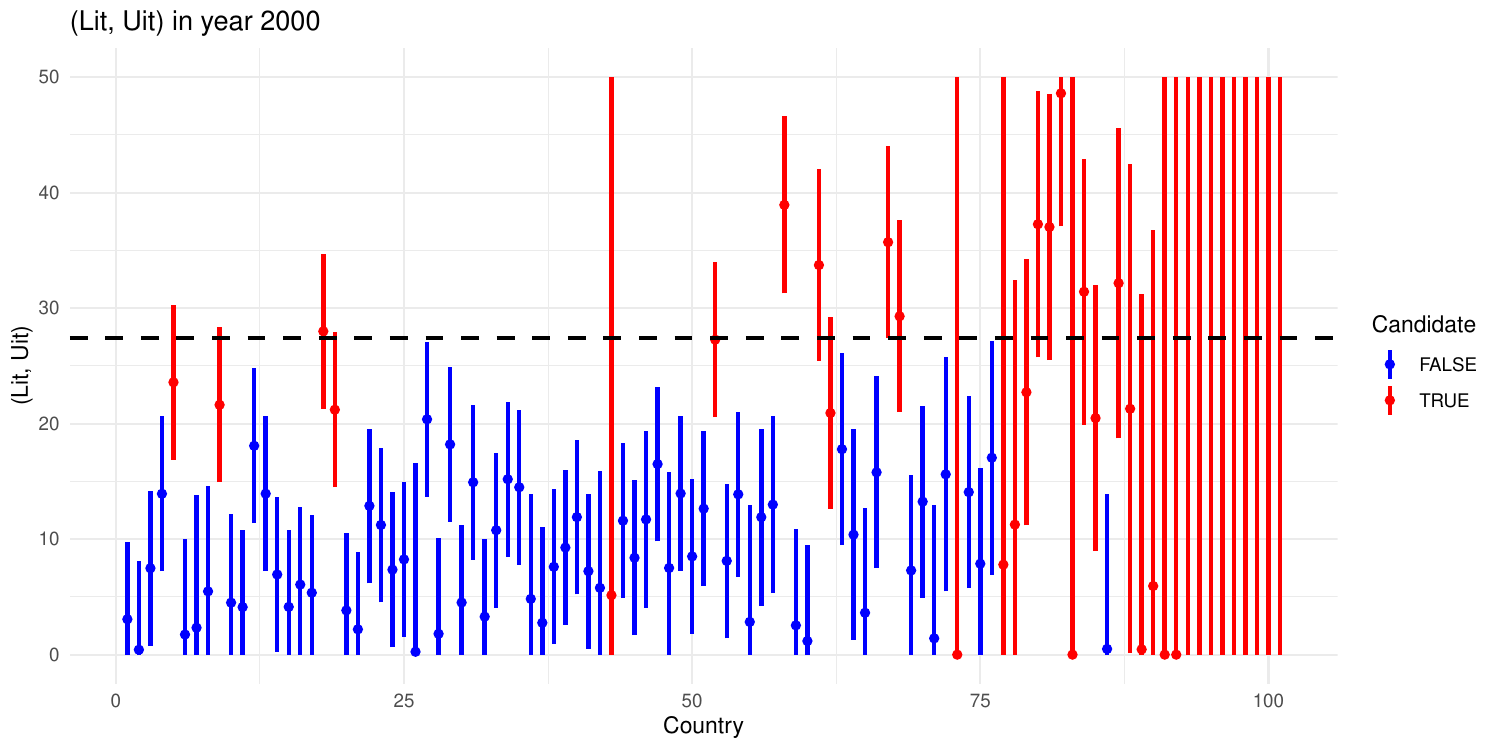}
    \caption{$(L_{it}(\alpha_{km}),U_{it}(\alpha_{km}))$ in year 2000.}
    \label{fig:suicide_ci_2000}
\end{figure}

\begin{figure}[H]
    \centering
    \includegraphics[width=1\linewidth]{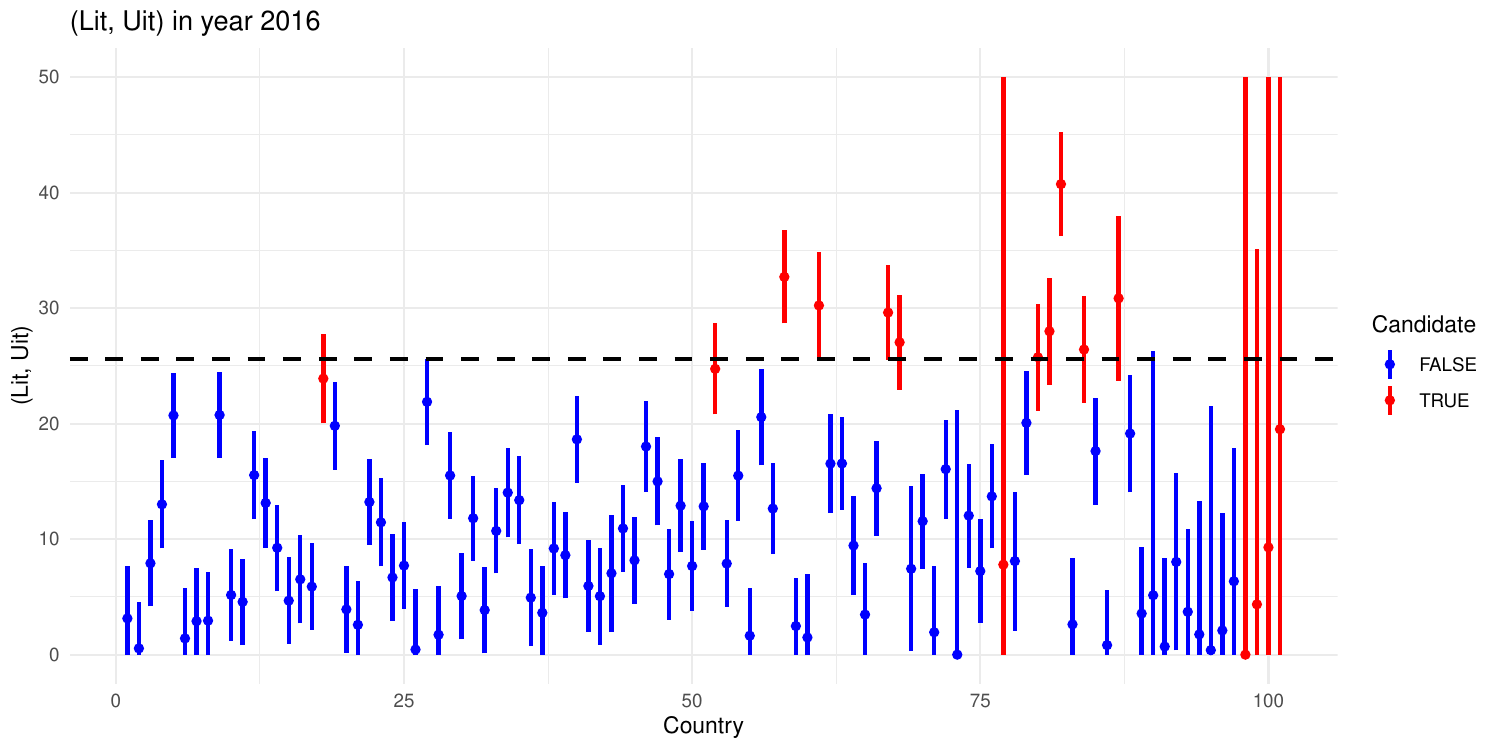}
    \caption{$(L_{it}(\alpha_{km}),U_{it}(\alpha_{km}))$ in year 2016.}
    \label{fig:suicide_ci_2016}
\end{figure}

\begin{figure}[H]
    \centering
    \includegraphics[width=1\linewidth]{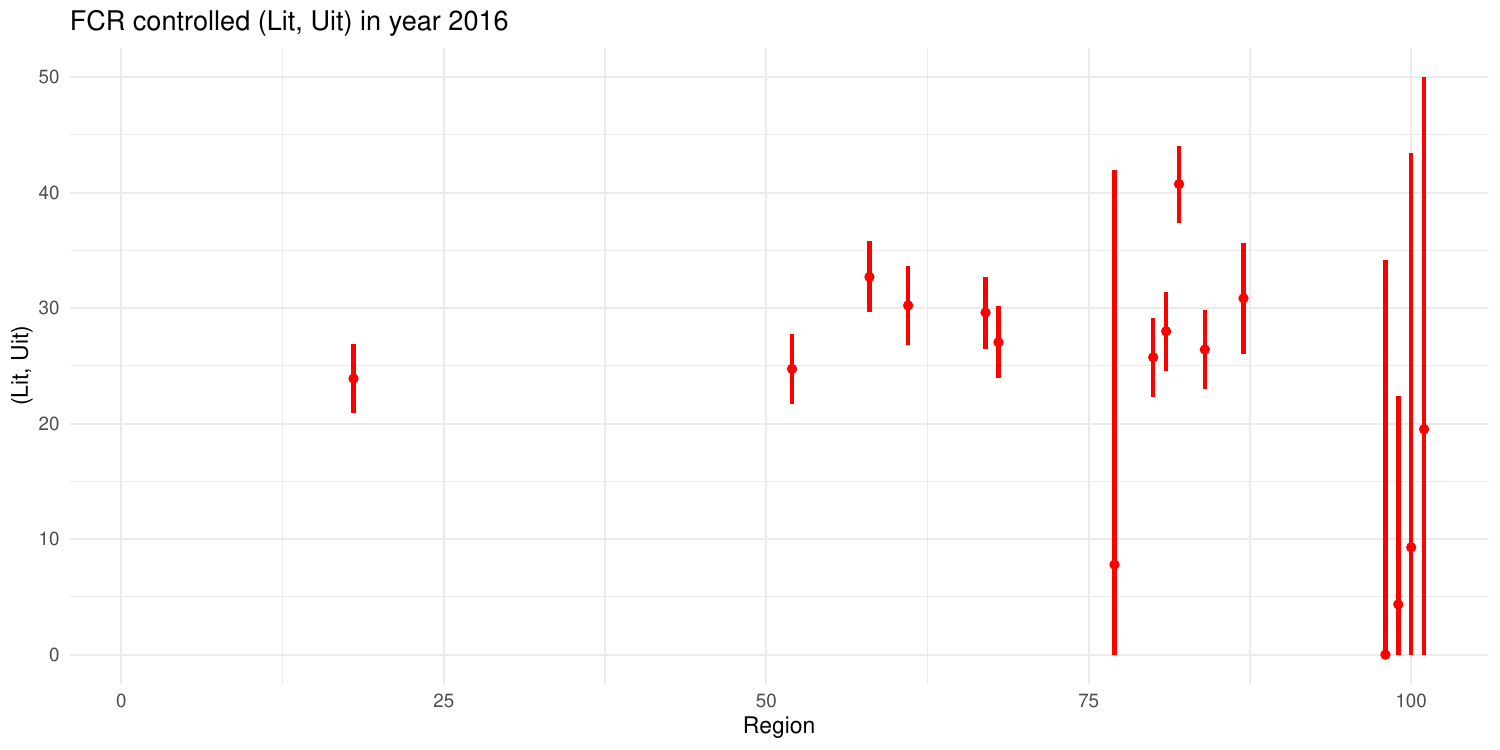}
    \caption{$(L_{it}(\tilde{\alpha}),U_{it}(\tilde{\alpha}))$ in year 2016.}
    \label{fig:suicide_FCR_2016}
\end{figure}

For the target $\theta_i$, the statistics are given by
\begin{align}
    L_{iT}(\alpha)=\bar{X}_{iT}-\frac{1}{\lambda T_i}\log\frac{1}{\alpha}-\frac{\sigma^2\lambda}{2}~~\text{and}~~U_{iT}(\alpha)=\bar{X}_{iT}+\frac{1}{\lambda T_i}\log\frac{1}{\alpha}+\frac{\sigma^2\lambda}{2},
\end{align}
where $L_{iT}(\alpha)=-\infty$ and $U_{iT}(\alpha)=\infty$ if $T_i=0$. As detailed in Section \ref{sec:supp} of Supplementary Material, we can verify that they satisfy \eqref{eq:bounded}. We set $\lambda$ to be the same for all $T$, which do not make $U_{it}(\alpha)$ and $L_{it}(\alpha)$ converge to $\theta_i$. In this case, Condition \ref{con:lln} is not satisfied, but it is not a big issue as sample sizes are small. As discussed in Section \ref{sec:discussion_lambda}, there is no reason to make $\lambda_{it}\xrightarrow{}0$ as $t\to\infty$. Moreover, given $T_i$, the length of $(L_{iT}(\alpha_{km}),U_{iT}(\alpha_{km}))$ becomes smallest when $\lambda=\sqrt{2\log(1/\alpha_{km})/(T_i\sigma^2)}$, which is about $0.15$ if $T_i=32$. However, since $T_i$ is not $\cF_{T-1}$-measurable, we cannot use this $\lambda$. To make $\lambda$ be $\cF_{T-1}$-measurable for each $T$, we set $\lambda=0.15$, leading to tighter intervals. The results are shown in Figures \ref{fig:suicide_1985}--\ref{fig:suicide_FCR_2016}.


In Figures \ref{fig:suicide_1985}--\ref{fig:suicide_2016}, the screened countries $i\in\hat{\cS}_T$ are colored red. We can see that the Nordic countries seem to have high suicide rates. Other than the Nordic countries, some countries are detected in Figure \ref{fig:suicide_2016} (e.g., Cape Verde, Dominica, Macao, Oman). However, they might have been selected because, as shown in Table \ref{table:suicide} in Section \ref{sec:table} of Supplementary Material, their sample sizes are too small to screen them out. 
Figures \ref{fig:suicide_ci_1985}--\ref{fig:suicide_ci_2016} show the intervals, $(L_{iT}(\alpha_{km}),U_{iT}(\alpha_{km}))$, where the dots represent $\bar{X}_{iT}$ and the black dashed line indicates $L_T^{(m)}(\alpha_{km})$. The red and blue colors mean $i\in\hat{\cS}_T$ and $i\notin\hat{\cS}_T$, respectively. In Figure \ref{fig:suicide_ci_2000}, some regions do not have points because $T_i=0$ in year 2000. In comparison between Figures \ref{fig:suicide_ci_2000} and \ref{fig:suicide_ci_2016}, we can see that each interval shrinks in $T$.
Finally, Figure \ref{fig:suicide_FCR_2016} represents the FCR-controlled interval $(L_{iT}(\tilde{\alpha}),U_{iT}(\tilde{\alpha}))$ for each $i\in\hat{\cS}_T$ in year 2016. Since $\tilde{\alpha}\geq\alpha_{km}$, we can observe that $(L_{iT}(\tilde{\alpha}),U_{iT}(\tilde{\alpha}))$ are narrower than $(L_{iT}(\alpha_{km}),U_{iT}(\alpha_{km}))$.

\section{Conclusion}\label{sec:conclusion}

In this paper, we have proposed two novel procedures: Sequential Correct Screening (SCS), a variable screening method, and Post-Screening Inference (PSI), a follow-up procedure. We have shown that SCS is an anytime-valid method designed to return a subset of variables that, with high probability, always contains the true top-$m$ variables. Furthermore, PSI provides a way to construct confidence intervals for the parameters selected by SCS while controlling the stopped False Coverage Rate (sFCR). The validity of proposed procedures is supported by the theory, simulation studies, and a real data example.

We end this chapter by pointing out some important directions for future studies. First, Property \ref{property} has a clear interpretation in terms of the Jaccard index (JAC), a similarity measure of two sets. Specifically, the JAC between sets $A$ and $B$ is defined as $\text{JAC}(A,B)=|A\cap B|/|A\cup B|$. This takes its values in $[0,1]$; in particular, it is equal to one if and only if $A=B$. Property \ref{property} entails that $\text{JAC}(\cS,\hat{\cS}_T)=|\cS|/|\hat{\cS}_T|$ and $\text{JAC}(\cS,\hat{\cS}_T)\to1$ with probability at least $1-\alpha$ as $T\to \infty$. Pursuing relationships to other metrics may influence related fields such as medical image segmentation \citep{bertels2019optimizing}. 
Second, Our procedures are valid under arbitrary dependence. If the dependence structure is known, it may be interesting to develop more powerful methodologies that adapt such dependence. 
Third, we have not considered an optimal sampling strategy, which is an important direction with many practical implications. For example, let $\tau=\min\{t\in\bbN:|\hat{\cS}_t|=m\}$ and make $\E[\tau]$ as small as possible. 
Fourth, in our asymptotic framework, the sample size in the time direction diverges while the number of parameters is fixed. Investigating the regime where the number of parameters varies across time is also important. In particular, \cite{kim2025locally} proposed a dimension-agnostic method to detect $1$-promising of means, which is valid without knowledge of the asymptotic regime. Their work may provide insights for this direction. 
Fifth, our SCS outputs subsets that can be larger than $m$ to cope with the uncertainty of a small sample size. A similar idea is found in \cite{liang2025assumption}, which also provides a subset of size greater than or equal to $m$. However, their primary focus is on the procedure's stability to small perturbations in the data, a topic that is also interesting to investigate for our method.

\section*{Supplementary Material}
Section A: Proofs, 
Section B: Table for Section \ref{sec:applications}, 
Section C: Justification of the Statistics in Section \ref{sec:applications},
Section D: Another Empirical Application.

\section*{Acknowledgments}
This work was supported by JSPS KAKENHI Grant Numbers JP25K00625, JP24K02904, by Grant-in-Aid for JSPS Fellows Grant Number JP25KJ1302, and by Nomura Foundation's Research Grant for Frontier of Financial and Capital Markets. The authors report there are no competing interests to declare.

\bibliographystyle{chicago}
\bibliography{ref}

\newpage
\appendix
\setcounter{page}{1}
\setcounter{section}{0}
\renewcommand{\theequation}{A.\arabic{equation}}
\setcounter{equation}{0}

\setcounter{table}{0}
\renewcommand{\thetable}{\thesection\arabic{table}}
\setcounter{figure}{0}
\renewcommand{\thefigure}{\thesection\arabic{figure}}

\begin{center}
{\Large Supplementary Material for} \\[7mm]
\textbf{\Large Sequential Correct Screening and \\[3mm]
    Post-Screening Inference} \\[10mm]
\textsc{\large Masaki Toyoda$^*$} {\large and} \textsc{\large Yoshimasa Uematsu$^\dagger$} \\[5mm]
*\textit{\large Department of Economics, Hitotsubashi University} \\[1mm]
$\dagger$\textit{\large Department of Social Data Science, Hitotsubashi University}
\end{center}

\section{Proofs}

\subsection{Proof of Theorem \ref{thm:correct}}\label{proofthm:correct}
\begin{proof}
    Recall that $\cS=[m]$, $\cS^c=[k]\backslash[m]$, $\alpha_{km}=\alpha/\{2m(k-m)\}$. We prove (a)--(c). First, (a) immediately holds by the procedure.
    
    Prove (b). It is sufficient to bound the probability of the complementary event by $\alpha$. First, the union bound and construction of $\hat\cS_T$ yield
    \begin{align}
        \Pro\left(\bigcup_{i=1}^{m}\{\exists T\in\bbN:i\notin\hat{\cS}_T\}\right)
        \leq\sum_{i=1}^{m}\Pro\left(\exists T\in\bbN:U_{iT}(\alpha_{km})<L_T^{(m)}(\alpha_{km})\right).
    \end{align}
Here we have $|\cS|=m$ by Condition \ref{con:promising} and $|\{i:L_{iT}(\alpha_{km})\geq L_T^{(m)}(\alpha_{km})\}|=m$ by the definition of $L_T^{(m)}(\alpha_{km})$. Thus, for any fixed $T\in\bbN$, if there exists some $i\in\cS$ such that $U_{iT}(\alpha_{km})<L_T^{(m)}(\alpha_{km})$, then there exists another $j\notin\cS$ that satisfies $U_{iT}(\alpha_{km})< L_{jT}(\alpha_{km})$. Therefore, for every $i\in[m]$, the probability is bounded by
    \begin{align}
        &\Pro\left(\exists T\in\bbN:U_{iT}(\alpha_{km})<L_T^{(m)}(\alpha_{km})\right)\\
        &\leq\Pro\left(\bigcup_{j\in \cS^c}\{\exists T\in\bbN:U_{iT}(\alpha_{km})<L_{jT}(\alpha_{km})\}\right)\\
        &\leq\sum_{j=m+1}^{k}\Pro\left(\exists T\in\bbN:U_{iT}(\alpha_{km})<L_{jT}(\alpha_{km})\right),
    \end{align}
    where the second inequality is due to the union bound.

For any $i\in\cS$ and $j\in\cS^c$, we have $\theta_i>\theta_j$ by Condition \ref{con:promising}. In this case, for any fixed $T\in\bbN$, $\theta_i<U_{iT}(\alpha_{km})$ and $\theta_j>L_{jT}(\alpha_{km})$ imply $U_{iT}(\alpha_{km})\geq L_{jT}(\alpha_{km})$. This contraposition entails that
    \begin{align}
        &\Pro\left(\exists T\in\bbN:U_{iT}(\alpha_{km})<L_{jT}(\alpha_{km})\right)\\
        &\leq \Pro\left(\exists T\in\bbN:\theta_i\geq U_{iT}(\alpha_{km})~\text{or}~\theta_j\leq L_{jT}(\alpha_{km}) \right)\\
        &\leq\Pro \left(\exists T\in\bbN:\theta_i\geq U_{iT}(\alpha_{km}) \right)+\Pro \left(\exists T:\theta_j\leq L_{jT}(\alpha_{km})\right)\\
        &\leq 2\alpha_{km}
\end{align}
    where the third inequality follows from \eqref{eq:bounded}. Recall $\alpha_{km}=\alpha/\{2m(k-m)\}$. Combining the results obtained so far gives
    \begin{align}   
       \Pro\left(\bigcup_{i=1}^{m}\{\exists T\in\bbN:i\notin\hat{\cS}_T\}\right) 
       \leq \sum_{i=1}^m\sum_{j=m+1}^k \frac{\alpha}{m(k-m)}
       =\alpha.
    \end{align}
This completes the proof of (b).

Prove (c). We begin with reducing the event:
\begin{align}
\left\{\cS=\hat{\cS}_T~\text{eventually}\right\}
\supset A \cap B \cap C,
\end{align}
where we have put 
$A= \{\forall T\in\bbN:\hat{\cS}_T\subset\hat{\cS}_{T-1}\}$, 
$B= \{\forall T\in\bbN:\cS\subset\hat{\cS}_T\}$, and 
$C= \{\exists T\in\bbN:\cS^c\cap\hat{\cS}_T=\emptyset\}$. Recall that Property \ref{property}(i) and (ii) establish $\Pro(A)=1$ and $\Pro(B)\geq 1-\alpha$, respectively. Thus, we have
\begin{align}
\Pro \left(\cS=\hat{\cS}_T~\text{eventually}\right)
&\geq \Pro(A\cap B\cap C) 
= \Pro(C\mid A\cap B) \Pro(A\cap B) \\
&\geq \Pro(C\mid A\cap B)(1-\alpha) .
\end{align}
Focusing on the conditioning on $A$, we have 
\begin{align}
\Pro(C\mid A\cap B) 
&= \Pro \left(\exists T\in\bbN:\bigcap_{i\in\cS^c}\left\{i \in\hat{\cS}_T^c\right\}\mid A\cap B\right) \\
&= \Pro \left(\bigcap_{i\in\cS^c}\left\{\exists T\in\bbN:i\in\hat{\cS}_T^c\right\}\mid A\cap B\right) \\
&\geq \sum_{i\in\cS^c} \Pro \left(\exists T\in\bbN:i\in\hat{\cS}_T^c\mid A\cap B\right) - (|\cS^c|-1),
\end{align}
where the inequality is due to an application of Bonferroni's inequality. Thus the proof of (c) completes if we prove that the last probability is equal to one for every $i\in\cS^c$. 

Fix any $i\in\cS^c$ hereafter. We obtain
\begin{align}
\Pro \left(\exists T\in\bbN:i\in\hat{\cS}_T^c\mid A\cap B\right) 
=\Pro \left(\exists T\in\bbN:U_{iT}(\alpha_{km})<L_T^{(m)}(\alpha_{km})\mid A\cap B\right).
\end{align}
Recall that $L_T^{(m)}(\alpha_{km})$ is the $m$-th largest order statistic among $\{L_{iT}(\alpha_{km}):i\in\hat\cS_{T-1}\}$. On the other hand, since $|\cS|=m$ by Condition \ref{con:promising}, $\min_{j\in\cS}L_{jT}(\alpha_{km})$ is also the $m$-th largest order statistic, but among the set $\{L_{iT}(\alpha_{km}):i\in\cS\}$. This implies that $\{\forall T\in\bbN:L_T^{(m)}(\alpha_{km})\geq\min_{j\in\cS}L_{jT}(\alpha_{km})\}$ conditional on $B$. Hence, we have
\begin{align}
 &\Pro \left(\exists T\in\bbN:U_{iT}(\alpha_{km})<L_T^{(m)}(\alpha_{km})\mid A\cap B \right)\\
&\geq\Pro\left(\exists T\in\bbN:U_{iT}(\alpha_{km})<\min_{j\in\cS}L_{jT}(\alpha_{km})\mid A\cap B\right)\\
&=1-\Pro\left(\forall T\in \bbN: M_{iT} \leq 0 \mid A\cap B\right),
\end{align}
where $M_{iT}=\min_{j\in\cS}L_{jT}(\alpha_{km})-U_{iT}(\alpha_{km})$. 
For all $T\in\bbN$, we have
\begin{align}
\Pro\left(\forall T\in \bbN: M_{iT} \leq 0 \mid A\cap B\right)
\leq \Pro\left(M_{iT} \leq 0 \mid A\cap B\right) 
\leq \frac{\Pro\left( M_{iT} \leq 0\right)}{1-\alpha}. \label{proof:thm1:001}
\end{align}
Condition \ref{con:lln} with the continuous mapping theorem yields $M_{iT}\xrightarrow{p}\mu_i:=\min_{j\in\cS}\theta_j-\theta_i$. Since $\mu_i>0$ by Condition \ref{con:promising}, this further implies that $\Pro(M_{iT}>0)\to 1$. 
Therefore, taking the limit of \eqref{proof:thm1:001} entails that $\Pro\left(\forall T\in \bbN: M_{iT} \leq 0 \mid A\cap B\right)=0$. This verifies the desired result, and thus completes the proof of (c).
\end{proof}

\subsection{Proof of Theorem \ref{thm:FCR}}\label{proofthm:FCR}
\begin{proof}
The proof is similar to \citesuppl{xu2024post}. 
Noting that $\tilde{\alpha}=\alpha|\hat{\cS}_\tau|/(2k)$ by Procedure \ref{proc2} and $\hat{\cS}_\tau\subset [k]$ a.s., we have
 \begin{align}
\text{sFCR}(\tilde{\alpha})&=\E\left[\sum_{i\in\hat{\cS}_\tau}\frac{\mathbb{I}\{\theta_i\notin (L_{i\tau}(\tilde{\alpha}),U_{i\tau}(\tilde{\alpha}))\}}{|\hat{\cS}_\tau|}\right]\\
&\leq \frac{\alpha}{2k}\sum_{i=1}^k\E\left[\frac{\mathbb{I}\{\theta_i\notin (L_{i\tau}(\tilde{\alpha}),U_{i\tau}(\tilde{\alpha}))\}}{\tilde{\alpha}}\right]. 
\end{align}
Condition \ref{con:posthoce} yields
\begin{align}
&\mathbb{I}\{\theta_i\notin (L_{i\tau}(\tilde{\alpha}),U_{i\tau}(\tilde{\alpha}))\}
=\mathbb{I}\{\theta_i\leq L_{i\tau}(\tilde{\alpha})~\text{or}~\theta_i\geq U_{i\tau}(\tilde{\alpha})\}\\
&\leq\mathbb{I}\left\{E_{i\tau}^-(\tilde{\alpha})\geq\frac{1}{\tilde{\alpha}}~\text{or}~E_{i\tau}^+(\tilde{\alpha})\geq\frac{1}{\tilde{\alpha}}\right\}
\leq\mathbb{I}\left\{\tilde{\alpha}(E_{i\tau}^-+E_{i\tau}^+) \geq 1\right\} \\
&\leq \tilde{\alpha}(E_{i\tau}^-+E_{i\tau}^+),
\end{align}
where the last inequality is true since the inequality $\mathbb{I}\{x\geq1\}\leq x$ holds for any $x\geq0$. Thus, using $\E [E_{i\tau}^-]+ \E [E_{i\tau}^+] \leq 2$ by Condition \ref{con:posthoce}, we obtain
\begin{align}
\text{sFCR}(\tilde{\alpha})
\leq \frac{\alpha}{2k}\sum_{i=1}^k\E\left[\frac{\tilde{\alpha}(E_{i\tau}^-+E_{i\tau}^+)}{\tilde{\alpha}}\right]\leq \alpha. 
\end{align}
This completes the proof.
\end{proof}

\subsection{Proof of Theorem \ref{thm:FCR1}}\label{proofthm:FCR1}
\begin{proof}
Recall that $|\hat{\cS}_\tau|\geq m$ a.s.\ by Procedure \ref{proc1}. For any $\alpha\in(0,1)$, we have
\begin{align}
\text{sFCR}(\alpha)&=\E\left[\sum_{i\in\hat{\cS}_\tau}\frac{\mathbb{I}\{\theta_i\notin (L_{i\tau}(\alpha),U_{i\tau}(\alpha))\}}{|\hat{\cS}_\tau|}\right]\\
&\leq\frac{1}{m}\sum_{i=1}^{k}\Pro \left(\theta_i\notin (L_{i\tau}(\alpha),U_{i\tau}(\alpha))\right)\\
&\leq\frac{1}{m}\sum_{i=1}^{k}\left\{\Pro(\theta_i\leq L_{i\tau}(\alpha)+\Pro(\theta_i\geq U_{i\tau}(\alpha))\right\}.
\end{align}
Lemma 1 of \citesuppl{ramdas2020admissible} states that \eqref{eq:bounded} is equivalent to
\begin{align}
\Pro(\theta_i\leq L_{i\tau}(\alpha))\leq\alpha~~\text{and}~~\Pro(\theta_i\geq U_{i\tau}(\alpha))\leq\alpha
\end{align}
for any $i\in[k]$ and any $\tau$, which leads to
\begin{align}
\Pro(\theta_i\leq L_{i\tau}(\alpha))+\Pro(\theta_i\geq U_{i\tau}(\alpha))
\leq 2\alpha.
\end{align}
Therefore, we conclude $\text{sFCR}(\alpha)\leq 2k\alpha/m$. 
Substituting $\alpha$ for $\alpha^\textsf{B}=m\alpha/(2k)$ yields the desired result.
\end{proof}

\subsection{Proof of Theorem \ref{thm:interval}}\label{proofthm:interval}
\begin{proof}
Prove (a).  If $m=1$, $k\geq2$, and $|\hat{\cS}_T|=1$, the statement is verified immediately. If $m=1$, $k\geq2$, and $|\hat{\cS}_T|\geq2$, then $\alpha^{\textsf{B}}<\alpha_{km}\leq\tilde{\alpha}$ is equivalent to $1< k/(k-1)\leq|\hat{\cS}_T|$ and this is true.

Prove (b). If $2\leq m\leq k-2$ and $k\geq4$, then $\alpha^{\textsf{B}}\leq\tilde{\alpha}$ holds since $|\hat{\cS}_T|\geq2$. Thus, it is sufficient to show that $\alpha_{km}\leq\alpha^{\textsf{B}}$, which is equivalent to $k\geq m^2/(m-1)$. Since $m^2/(m-1)$ is increasing in $m\geq2$, noting that $k\geq(k-2)^2/(k-3)\geq0$ verifies the statement.

Prove (c). If $m=k-1$ and $k\geq3$, then $\alpha^{\textsf{B}}<\alpha_{km}\leq\tilde{\alpha}$ is equivalent to $k<(k-1)^2$ and this is true.
\end{proof}

\section{Table for Section \ref{sec:applications}}\label{sec:table}

\begin{longtable}{llllll}
\caption{List of country names and $T_i$.}\label{table:suicide}\\
\hline
\footnotesize
\textbf{No.} & \textbf{Country} & \textbf{$T_i$} & \textbf{No.} & \textbf{Country} & \textbf{$T_i$} \\
\hline
\endhead
1  & Albania                      & 22  & 2  & Antigua and Barbuda          & 27  \\
3  & Argentina                    & 31  & 4  & Australia                    & 30  \\
5  & Austria                      & 32  & 6  & Bahamas                      & 23  \\
7  & Bahrain                      & 21  & 8  & Barbados                     & 25  \\
9  & Belgium                      & 31  & 10 & Belize                       & 28  \\
11 & Brazil                       & 31  & 12 & Bulgaria                     & 30  \\
13 & Canada                       & 29  & 14 & Chile                        & 31  \\
15 & Colombia                     & 31  & 16 & Costa Rica                   & 30  \\
17 & Ecuador                      & 31  & 18 & Finland                      & 29  \\
19 & France                       & 30  & 20 & Greece                       & 31  \\
21 & Guatemala                    & 30  & 22 & Iceland                      & 32  \\
23 & Ireland                      & 30  & 24 & Israel                       & 31  \\
25 & Italy                        & 31  & 26 & Jamaica                      & 17  \\
27 & Japan                        & 31  & 28 & Kuwait                       & 25  \\
29 & Luxembourg                   & 31  & 30 & Malta                        & 31  \\
31 & Mauritius                    & 32  & 32 & Mexico                       & 31  \\
33 & Netherlands                  & 32  & 34 & New Zealand                  & 29  \\
35 & Norway                       & 30  & 36 & Panama                       & 25  \\
37 & Paraguay                     & 27  & 38 & Portugal                     & 27  \\
39 & Puerto Rico                  & 31  & 40 & Republic of Korea            & 31  \\
41 & Saint Lucia                  & 28  & 42 & Saint Vincent and Grenadines & 25  \\
43 & Seychelles                   & 18  & 44 & Singapore                    & 31  \\
45 & Spain                        & 31  & 46 & Suriname                     & 28  \\
47 & Sweden                       & 30  & 48 & Thailand                     & 28  \\
49 & Trinidad and Tobago          & 27  & 50 & Turkmenistan                 & 29  \\
51 & USA                          & 31  & 52 & Ukraine                      & 28  \\
53 & United Kingdom               & 31  & 54 & Uruguay                      & 28  \\
55 & Grenada                      & 26  & 56 & Guyana                       & 25  \\
57 & Romania                      & 28  & 58 & Russia                       & 27  \\
59 & Armenia                      & 25  & 60 & Azerbaijan                   & 16  \\
61 & Belarus                      & 21  & 62 & Cuba                         & 24  \\
63 & Czech Republic               & 27  & 64 & El Salvador                  & 24  \\
65 & Georgia                      & 22  & 66 & Germany                      & 26  \\
67 & Hungary                      & 26  & 68 & Kazakhstan                   & 26  \\
69 & Kiribati                     & 11  & 70 & Kyrgyzstan                   & 26  \\
71 & Philippines                  & 15  & 72 & Poland                       & 24  \\
73 & Saint Kitts and Nevis        & 3   & 74 & Slovakia                     & 22  \\
75 & Uzbekistan                   & 22  & 76 & Denmark                      & 22  \\
77 & Macao                        & 1   & 78 & Aruba                        & 14  \\
79 & Croatia                      & 22  & 80 & Estonia                      & 21  \\
81 & Latvia                       & 21  & 82 & Lithuania                    & 22  \\
83 & Qatar                        & 15  & 84 & Slovenia                     & 21  \\
85 & Switzerland                  & 21  & 86 & South Africa                 & 20  \\
87 & Sri Lanka                    & 11  & 88 & Serbia                       & 18  \\
89 & Cyprus                       & 15  & 90 & San Marino                   & 3   \\
91 & Maldives                     & 10  & 92 & Montenegro                   & 10  \\
93 & Fiji                         & 11  & 94 & United Arab Emirates         & 6   \\
95 & Oman                         & 3   & 96 & Turkey                       & 7   \\
97 & Nicaragua                    & 6   & 98 & Dominica                     & 1   \\
99 & Bosnia and Herzegovina       & 2   & 100 & Cape Verde                  & 1   \\
101 & Mongolia                    & 1   &    &                               &     \\
\hline
\end{longtable}

\section{Justification of the Statistics in Section \ref{sec:applications}}\label{sec:supp}
In this section, we verify that the statistics, $U_{iT}(\alpha)$ and $L_{iT}(\alpha)$, in Section \ref{sec:applications} satisfy \eqref{eq:bounded}. Recall that $X_{iT}$ are $\sigma^2$-subGaussian. Therefore, for each $\lambda\in\bbR$,
\begin{align}
    \E\left[\exp\left\{\lambda(-X_{iT}+\theta_i)\right\}\mid X_{i1},\dots,X_{i,T-1}\right]
    &\leq\exp\left\{\frac{\sigma^2\lambda^2}{2}\right\}
   ~~\text{and}\label{eq:supplement}\\
    \E\left[\exp\left\{\lambda(X_{iT}-\theta_i)\right\}\mid X_{i1},\dots,X_{i,T-1}\right]
    &\leq\exp\left\{\frac{\sigma^2\lambda^2}{2}\right\}.
\end{align}
Let
\begin{align}
    M_{iT}^+&=\prod_{t=1}^{T}\exp\left\{\lambda(-X_{it}+\theta_i)\mathbb{I}\{\text{$X_{it}$ is observed}\}-\frac{\sigma^2\lambda^2\mathbb{I}\{\text{$X_{it}$ is observed}\}}{2}\right\}~~,\\
    M_{iT}^-&=\prod_{t=1}^{T}\exp\left\{\lambda(X_{it}-\theta_i)\mathbb{I}\{\text{$X_{it}$ is observed}\}-\frac{\sigma^2\lambda^2\mathbb{I}\{\text{$X_{it}$ is observed}\}}{2}\right\},
\end{align}
and define $\cF_T=\sigma(X_{it}\mathbb{I}\{\text{$X_{it}$ is observed}\},\mathbb{I}\{\text{$X_{it}$ is observed}\}:t\leq T)$. We will check that $(M_{iT}^+)_T$ and $(M_{iT}^-)_T$ are nonnegative supermartingales with respect to $\cF_T$. Note that
\begin{align}
    &\E\left[M_{iT}^+\mid\cF_{T-1}\right]\\
    &=M_{i,T-1}^+\E\left[\exp\left\{\lambda(-X_{iT}+\theta_i)\mathbb{I}\{\text{$X_{iT}$ is observed}\}-\frac{\sigma^2\lambda^2\mathbb{I}\{\text{$X_{iT}$ is observed}\}}{2}\right\}\mid\cF_{T-1}\right].
\end{align}
\begin{itemize}
    \item If $\mathbb{I}\{\text{$X_{iT}$ is observed}\}=0$, we have
    \begin{align}
        \E\left[M_{iT}^+\mid\cF_{T-1}\right]
        =M_{i,T-1}^+\E[1\mid\cF_{T-1}]=M_{i,T-1}^+.
    \end{align}
    \item If $\mathbb{I}\{\text{$X_{iT}$ is observed}\}=1$, we have
    \begin{align}
        \E\left[M_{iT}^+\mid\cF_{T-1}\right]
        &\leq M_{i,T-1}^+\E\left[\exp\left\{\lambda(-X_{iT}+\theta_i)-\sigma^2\lambda^2/2\right\}\mid\cF_{T-1}\right]\\
        &=M_{i,T-1}^+\E\left[\exp\left\{\lambda(-X_{iT}+\theta_i)-\sigma^2\lambda^2/2\right\}\mid X_{i1},\dots,X_{i,T-1}\right]\\
        &\leq M_{i,T-1}^+
    \end{align}
    where the equality holds since we fix $\mathbb{I}\{\text{$X_{iT}$ is observed}\}=1$, and the last inequality follows from \eqref{eq:supplement}.
\end{itemize}
Therefore, we have checked that $(M_{iT}^+)_T$ is a nonnegative supermartingale with respect to $\cF_T$. Similarly, we can check that $(M_{iT}^-)_T$ is a nonnegative supermartingale with respect to $\cF_T$ as well. Moreover, we also have $\E[M_{i1}^+]\leq1$ and $\E[M_{i1}^-]\leq1$. Consequently, Ville's inequality implies that
\begin{align}
    \Pro\left(\exists T\in\bbN: M_{iT}^+\geq1/\alpha\right)\leq\alpha~~\text{and}~~
    \Pro\left(\exists T\in\bbN: M_{iT}^-\geq1/\alpha\right)\leq\alpha.
\end{align}
Recall that $T_i=\sum_{t=1}^{T}\mathbb{I}\{\text{$X_{it}$ is observed}\}$ and $\lambda>0$ in Section \ref{sec:applications}. If $T_i\neq0$, we have
\begin{align}
    &\quad M_{iT}^+\geq1/\alpha\\
    &\Longleftrightarrow\lambda\sum_{t=1}^{T}(-X_{it}+\theta_i)\mathbb{I}\{\text{$X_{it}$ is observed}\}-\frac{\sigma^2\lambda^2}{2}\sum_{t=1}^{T}\mathbb{I}\{\text{$X_{it}$ is observed}\}\geq\log\frac{1}{\alpha}\\
    &\Longleftrightarrow-\lambda\sum_{t=1}^{T}X_{it}\mathbb{I}\{\text{$X_{it}$ is observed}\}+\lambda\theta_i T_i-\frac{\sigma^2\lambda^2}{2}T_i\geq\log\frac{1}{\alpha}\\
    &\Longleftrightarrow\theta_i\geq U_{iT}(\alpha).
\end{align}
If $T_i=0$, we have $M_{iT}^+=1<1/\alpha$. Therefore, we can conclude that
\begin{align}
    \Pro(\exists T\in\bbN:\theta_i\geq U_{iT}(\alpha))
    &=\Pro(\exists T\in\bbN:\theta_i\geq U_{iT}(\alpha),~T_i>0)\\
    &=\Pro\left(\exists T\in\bbN: M_{iT}^+\geq1/\alpha\right)\\
    &\leq\alpha.
\end{align} 
Similarly, we can also conclude that $\Pro(\exists T\in\bbN:\theta_i\leq L_{iT}(\alpha))\leq\alpha$.

\section{Another Empirical Application}\label{sec:chocolate}
We apply our procedures to the chocolate bar ratings data, obtained from \url{https://www.kaggle.com/datasets/rtatman/chocolate-bar-ratings}. Each row records the flavor rating for a chocolate, the regions where the cocoa beans are grown, and the order in which the rating was entered into the database. Although there are data with the same order, we consider that these data are obtained simultaneously. The number of unique orders is $440$. Before the analysis, we deleted the rows where the regions are missing and combined the rows with essentially the same regions (specifically, we combined the rows corresponding to ``Venezuela/ Ghana'' and ``Venezuela, Ghana''). Then the number of unique regions becomes $98$.

\begin{figure}[H]
    \centering
    \includegraphics[width=1\linewidth]{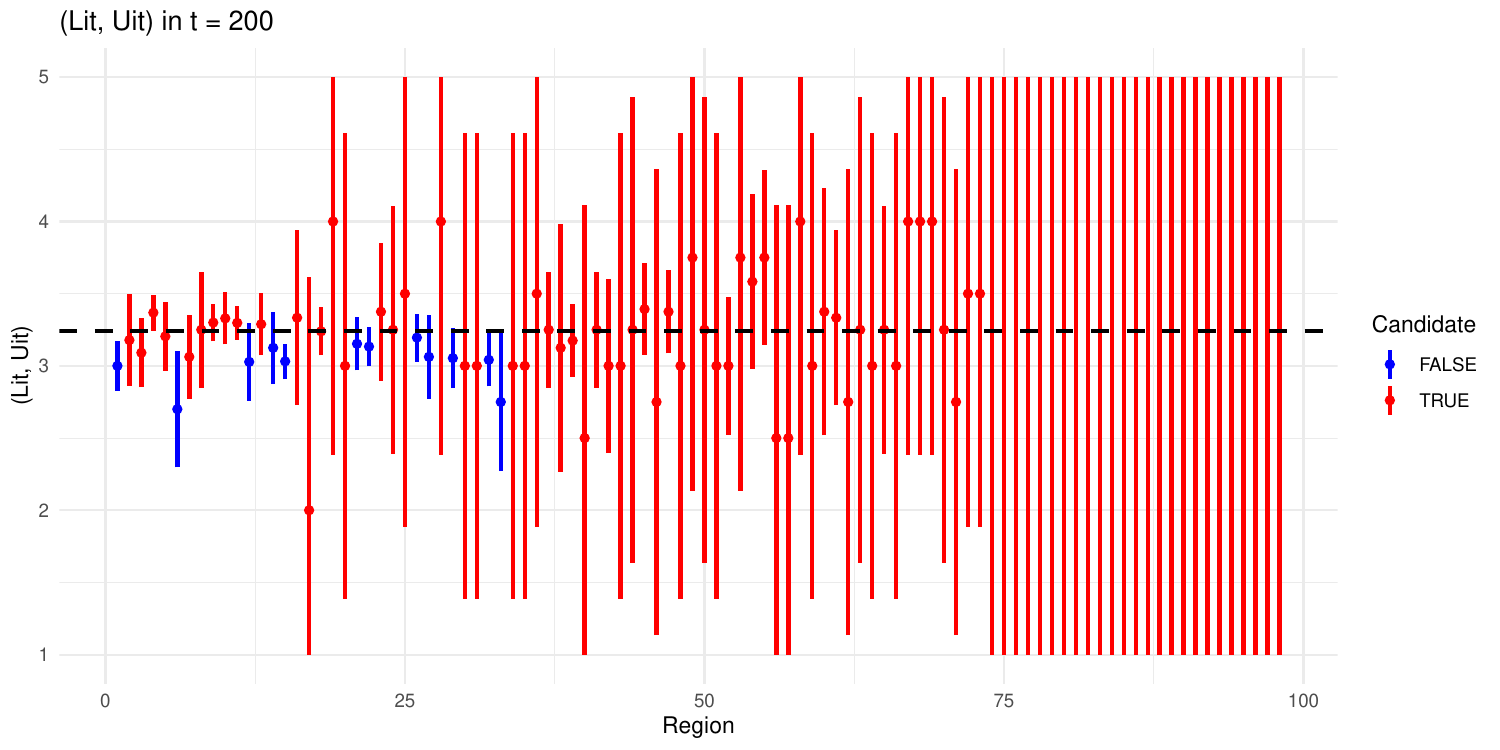}
    \caption{$(L_{it}(\alpha_{km}),U_{it}(\alpha_{km}))$ with $t=200$.}
    \label{fig:200_chocolate}
\end{figure}

\begin{figure}[H]
    \centering
    \includegraphics[width=1\linewidth]{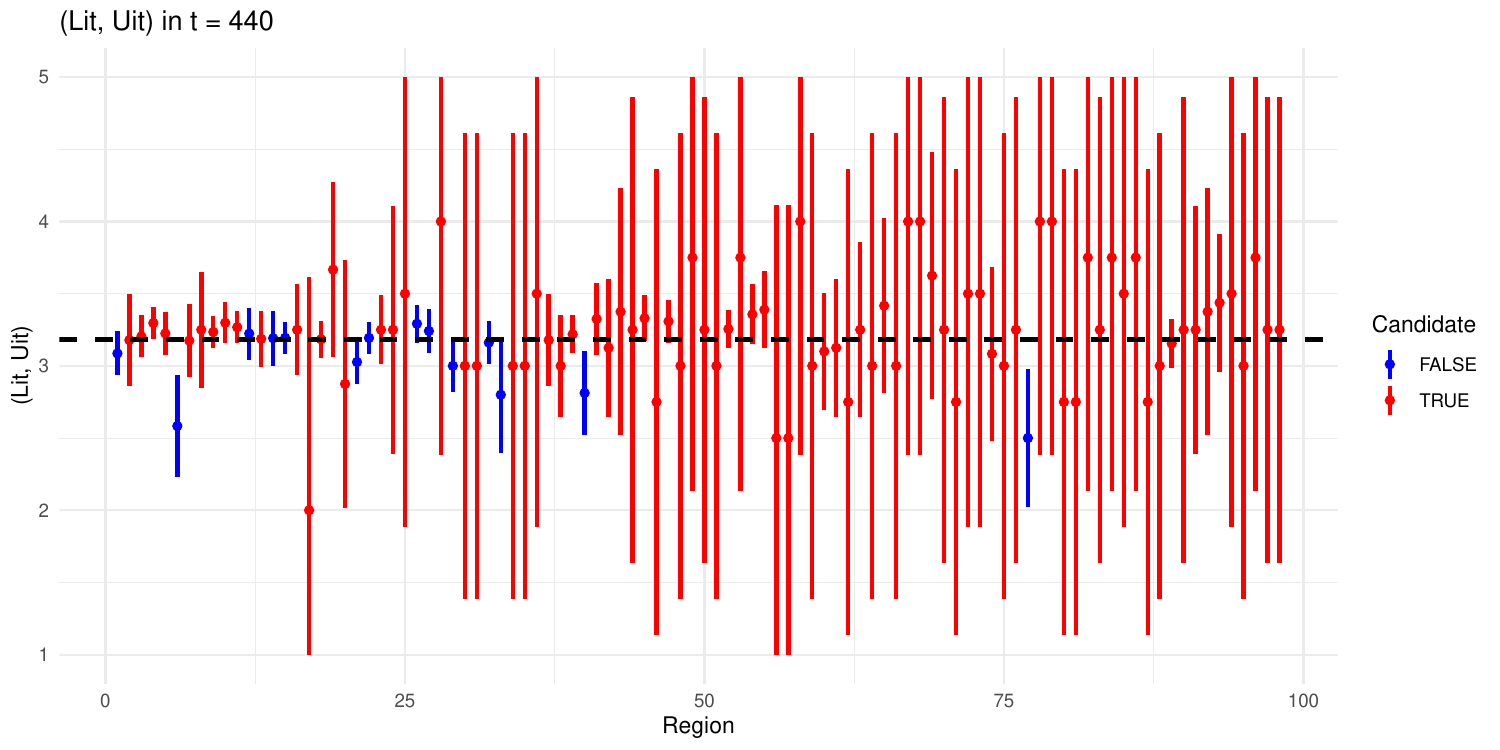}
    \caption{$(L_{it}(\alpha_{km}),U_{it}(\alpha_{km}))$ with $t=440$.}
    \label{fig:440_chocolate}
\end{figure}

\begin{figure}[H]
    \centering
    \includegraphics[width=1\linewidth]{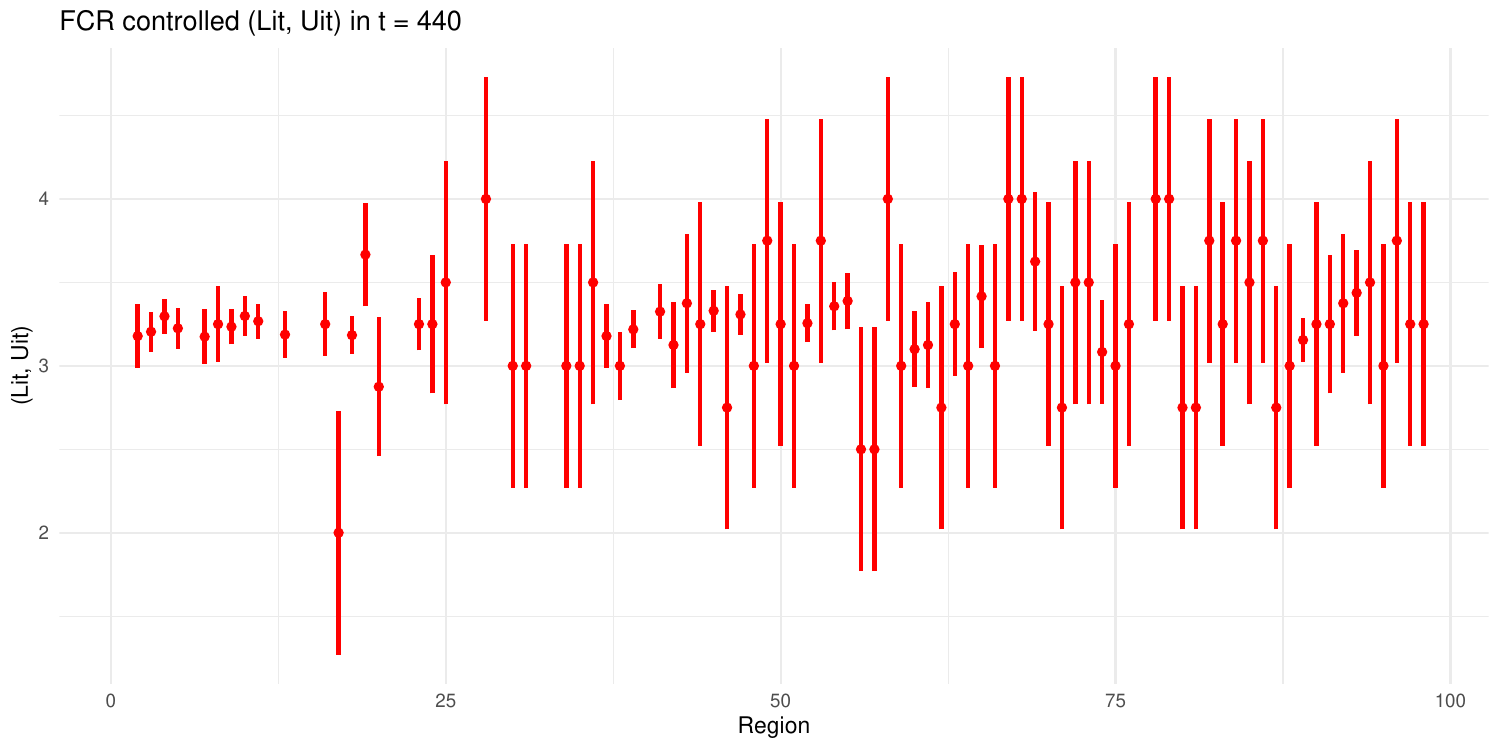}
    \caption{$(L_{it}(\tilde{\alpha}),U_{it}(\tilde{\alpha}))$ with $t=440$.}
    \label{fig:440_chocolate_FCR}
\end{figure}

Let $X_{iT}$ be the rating of a chocolate of region $i\in[98]$ and order $T\in[440]$, and suppose that $X_{i1},\dots,X_{i,440}$ are i.i.d.\ $\sigma^2$-subGaussian with $\theta_i=\E[X_{i1}]$. Although $X_{iT}$ are missing for some pairs $(i,T)$, we use only observed data up to the order $T$ to construct $L_{iT}(\alpha)$ and $U_{iT}(\alpha)$. Specifically, let $T_i=\sum_{t=1}^{T}\mathbb{I}\{\text{$X_{it}$ is observed}\}$ be the number of observations in $X_{i1},\dots,X_{iT}$, and let $\bar{X}_{iT}=(1/T_i)\sum_{t=1}^{T}X_{it}\mathbb{I}\{\text{$X_{it}$ is observed}\}$ be the sample mean of observed data among $X_{i1},\dots,X_{iT}$. As detailed in Section \ref{sec:supp}, we can check that
\begin{align}
    L_{iT}(\alpha)=\bar{X}_{iT}-\frac{1}{\lambda T_i}\log\frac{1}{\alpha}-\frac{\sigma^2\lambda}{2}~~\text{and}~~
    U_{iT}(\alpha)=\bar{X}_{iT}+\frac{1}{\lambda T_i}\log\frac{1}{\alpha}+\frac{\sigma^2\lambda}{2}
\end{align}
satisfy \eqref{eq:bounded} for any fixed $\lambda\in\bbR$ and $\alpha\in(0,1)$. We set $\alpha=0.1$, $\sigma=0.2$, $m=1$ and $\lambda=5$. Since $\lambda$ is same for all $T$, $U_{iT}(\alpha)$ and $L_{iT}(\alpha)$ do not converge to $\theta_i$ so that Condition \ref{con:lln} is not satisfied. However, they can be narrow intervals since the sample sizes are relatively small. The results are shown in Table \ref{table:chocolate}, Figures \ref{fig:200_chocolate}, \ref{fig:440_chocolate} and \ref{fig:440_chocolate_FCR}.

\begin{longtable}{llllll}
\caption{List of region names and $T_i$ at $T=440$.}\label{table:chocolate}\\
\hline
\textbf{No.} & \textbf{Region} & \textbf{$T_i$} & \textbf{No.} & \textbf{Region} & \textbf{$T_i$} \\
\hline
\endhead
1  & Ghana                        & 29  & 2  & Sao Tome \& Principe  & 7  \\
3  & Trinidad                     & 33  & 4  & Madagascar            & 117  \\
5  & Colombia                     & 30  & 6  & West Africa           & 6  \\
7  & Sao Tome                     & 10  & 8  & Carribean             & 5  \\
9  & Dominican Republic           & 114 & 10 & Papua New Guinea      & 36  \\
11 & Venezuela                    & 157 & 12 & Hawaii                & 19  \\
13 & Indonesia                    & 16  & 14 & Jamaica               & 17  \\
15 & Ecuador                      & 143 & 16 & Vanuatu               & 7  \\
17 & Ghana \& Madagascar          & 1   & 18 & Bolivia               & 53  \\
19 & South America                & 3   & 20 & Sri Lanka             & 2  \\
21 & Mexico                       & 28  & 22 & Peru                  & 128  \\
23 & Cuba                         & 11  & 24 & Venezuela, Ghana      & 2  \\
25 & Indonesia, Ghana             & 1   & 26 & Brazil                & 48  \\
27 & Tanzania                     & 30  & 28 & Venezuela, Java       & 1  \\
29 & Grenada                      & 18  & 30 & Carribean(DR/Jam/Tri) & 1  \\
31 & Venezuela, Carribean         & 1   & 32 & Costa Rica            & 31  \\
33 & Ivory Coast                  & 5   & 34 & Ghana, Panama, Ecuador & 1  \\
35 & Ven., Trinidad, Mad.         & 1   & 36 & Ven., Indonesia, Ecuad. & 1  \\
37 & Panama                       & 7   & 38 & St. Lucia             & 6  \\
39 & Nicaragua                    & 48  & 40 & Uganda                & 8  \\
41 & Congo                        & 10  & 42 & Philippines           & 4  \\
43 & Trinidad, Tobago             & 2   & 44 & Ven, Trinidad, Ecuador & 1  \\
45 & Guatemala                    & 25  & 46 & Cost Rica, Ven        & 1  \\
47 & Vietnam                      & 30  & 48 & Ghana, Domin. Rep     & 1  \\
49 & Dominican Rep., Bali         & 1   & 50 & Nigeria               & 1  \\
51 & Colombia, Ecuador            & 46  & 52 & Belize                & 1  \\
53 & DR, Ecuador, Peru            & 14  & 54 & Honduras              & 9  \\
55 & Haiti                        & 1   & 56 & Peru, Madagascar      & 1  \\
57 & Venezuela, Trinidad          & 1   & 58 & Ven, Bolivia, D.R.    & 1  \\
59 & South America, Africa        & 5   & 60 & Fiji                  & 4  \\
61 & India                        & 1   & 62 & Principe              & 3  \\
63 & Australia                    & 1   & 64 & Ecuador, Mad., PNG    & 1  \\
65 & Central and S. America       & 1   & 66 & Peru, Ecuador, Venezuela & 1  \\
67 & Dom. Rep., Madagascar        & 2   & 68 & Gre., PNG, Haw., Haiti, Mad & 1  \\
69 & Tobago                       & 1   & 70 & Peru, Ecuador         & 1  \\
71 & Venezuela, Dom. Rep.         & 1   & 72 & Peru, Mad., Dom. Rep. & 1  \\
73 & Mad., Java, PNG              & 1   & 74 & Liberia               & 3  \\
75 & Burma                        & 1   & 76 & Gabon                 & 1  \\
77 & Puerto Rico                  & 4   & 78 & Guat., D.R., Peru, Mad., PNG & 1  \\
79 & Peru, Dom. Rep               & 1   & 80 & Africa, Carribean, C. Am. & 1  \\
81 & Martinique                   & 1   & 82 & PNG, Vanuatu, Mad     & 1  \\
83 & Madagascar \& Ecuador        & 1   & 84 & Peru, Belize          & 1  \\
85 & Trinidad, Ecuador            & 1   & 86 & Venez,Africa,Brasil,Peru,Mex & 1  \\
87 & Trinidad-Tobago              & 21  & 88 & El Salvador           & 1  \\
89 & Domincan Republic            & 2   & 90 & Cameroon              & 2  \\
91 & Togo                         & 4   & 92 & Malaysia              & 1  \\
93 & Solomon Islands              & 1   & 94 & Samoa                 & 1  \\
95 & Peru(SMartin,Pangoa,nacional) & 1   & 96 & Ven.,Ecu.,Peru,Nic.   & 1  \\
97 & Suriname                     & 1   & 98 & Ecuador, Costa Rica   & 1  \\
\hline
\end{longtable}


In Figures \ref{fig:200_chocolate} and \ref{fig:440_chocolate}, each interval represents $(L_{iT}(\alpha_{km}),U_{iT}(\alpha_{km}))$, each point represents $\bar{X}_{iT}$, and the black dashed line represents $L_T^{(m)}(\alpha_{km})$. The red color means $i\in\hat{\cS}_T$ and the blue color means $i\notin\hat{\cS}_T$. In Figure \ref{fig:200_chocolate}, some regions do not have points because $T_i=0$ in $T=200$. By comparing Figures \ref{fig:200_chocolate} and \ref{fig:440_chocolate}, we can see that each interval shrinks in $T$. Note that $i\notin\hat{\cS}_T$ if $U_{iT}(\alpha_{km})$ has been smaller than $L_T^{(m)}(\alpha_{km})$ at least once. Therefore, it is not surprising that there are some blue intervals that cross the black line. Moreover, although it seems that $\hat{\cS}_{440}$ is not so small, it may be natural since many regions have very small sample sizes $t_i$ as shown in Table \ref{table:chocolate}.

Finally, Figure \ref{fig:440_chocolate_FCR} represents $(L_{iT}(\tilde{\alpha}),U_{iT}(\tilde{\alpha}))$ for each $i\in\hat{\cS}_T$ in $T=440$. Since $\tilde{\alpha}\geq\alpha_{km}$, we can observe that FCR controlling intervals $(L_{iT}(\tilde{\alpha}),U_{iT}(\tilde{\alpha}))$ are narrower than $(L_{iT}(\alpha_{km}),U_{iT}(\alpha_{km}))$.

\end{document}